\documentclass[pra,onecolumn,superscriptaddress,aps,5pt]{revtex4-1}
\usepackage{amsfonts}
\usepackage{amssymb,latexsym,amsmath}
\usepackage{graphicx}
\usepackage{dcolumn}
\usepackage{times}
\usepackage{subfigure}
\usepackage[toc,page,title,titletoc,header]{appendix}
\usepackage[usenames]{color}
\usepackage{ulem}

\newcommand{\sech}{\mathrm{sech}}

\newcommand{\intl}{\int_{-\infty}^{+\infty}}
\begin{document}

\title{Dark-bright Solitons and their Lattices in
% Homogeneous and Trapped
Atomic Bose-Einstein Condensates}

\author{D. Yan}
\affiliation{Department of Mathematics and Statistics, University of
Massachusetts, Amherst MA 01003-4515, USA}

\author{F. Tsitoura}
\affiliation{Department of Physics, University of Athens,
Panepistimiopolis, Zografos, Athens 15784, Greece}
%
%\author{V. Achilleos}
%\affiliation{Department of Physics, University of Athens,
%Panepistimiopolis, Zografos, Athens 15784, Greece}

\author{P. G. Kevrekidis}
\affiliation{Department of Mathematics and Statistics, University of
Massachusetts, Amherst MA 01003-4515, USA}
\author{D. J. Frantzeskakis}
\affiliation{Department of Physics, University of Athens,
Panepistimiopolis, Zografos, Athens 15784, Greece}

\begin{abstract}
In the present contribution, we explore a host of different
stationary states, namely dark-bright solitons and their lattices,
that arise in the context of multi-component atomic
Bose-Einstein condensates. The latter, are modeled by systems of coupled
%multi-component
Gross-Pitaevskii equations with
%under
general interaction (nonlinearity)
coefficients $g_{ij}$. It is found that in some particular
parameter ranges such solutions can be obtained in analytical
form, however, numerically they are computed as existing in a
far wider parametric range.
%We consider the states both without
%(homogeneous condensates) and with a parabolic trap and find that
%they can be sustained in both settings.
Many features of the solutions
under study, such as their analytical form without the trap or the
stability/dynamical properties of one dark-bright soliton even in
the presence of the trap are obtained analytically and corroborated
numerically. Additional features, such as the stability of
soliton lattice homogeneous states or their existence/stability
in the presence of the trap, are examined numerically.
\end{abstract}

\maketitle

\section{Introduction}

Dark-bright (DB) solitons
%solitary waves
constitute exact solutions of the completely integrable,
%two-component
defocusing, two-component Manakov model~\cite{manakov},
i.e., the vector
variant of the Nonlinear Schr{\"o}dinger equation~\cite{APT}.
These structures exist in the presence of equal
%self-defocusing (equal)
nonlinear interactions within and between components. As such, they can
be thought of as symbiotic structures, since the bright components
thereof would not be sustainable in defocusing
%such self-repulsive
settings, and only emerge because of the effective potential well created
by the dark soliton component through the inter-species interaction.

Taking advantage of the ratios of inter- and intra-species interactions
between Bose-condensed hyperfine spin states of
atomic $^{87}$Rb, being very proximal to unity,
dark-bright solitons were proposed as being experimentally
relevant in atomic Bose-Einstein condensates (BECs) already since 2001
%in the work of~
\cite{buschanglin}. However,
this possibility was at a somewhat
dormant stage until 2008, when the Hamburg group
%an experiment using phase-imprinting techniques~\cite{becker}
was able to produce experimentally such coherent structures
using phase-imprinting techniques~\cite{becker},
and to illustrate their robustness in $^{87}$Rb BECs.
%experiments.
The above mentioned as well as subsequent efforts revealed a number of exciting
characteristics of these nonlinear entities. For instance, it was
shown that DB solitary waves oscillate in a trap with a reduced
frequency in comparison to their dark single-component counterparts
due to the presence of the bright filling component~\cite{buschanglin,becker,pe1}.
Dark-bright soliton trains were created by inducing counterflow between
two miscible BECs past a critical velocity~\cite{pe2}.
Molecules of a few DB solitary waves were observed in related
experiments, and offered the seed for detailed investigations
of the interactions between DB solitons
%DB-DB soliton interaction~
\cite{pe2a,vpg1,vpg2}.
Furthermore, beating (in time) dark-dark solitons,
%solitary waves,
which turn out to be
%beating (in time)
SO(2) rotated versions of DB solitons were also predicted and observed
%found to be present
in experiments~\cite{pe3,pe4}, further adding to the richness
of this multi-component setting. Also, the interaction of
such states with potential barriers was experimentally explored~\cite{pejc}.
It should also be noted that
two-dimensional generalizations of these structures have been
considered, both in the context of dark-bright rings~\cite{jan_stock}
and in that of vortex-bright solitary waves~\cite{kody_prl,martina2}.

Our aim in the present work is to present a set of analytical
solutions and numerical results both for individual DB
%dark-bright
solitary waves and also for
%the context of
lattices of such waves, for arbitrary nonlinear
coefficients (within suitable bounds). This is relevant
for a number of reasons not only theoretically, but also experimentally.
On the one hand, not all atomic species have as nearly equal
inter- and intra-species interaction scattering length, as is the
case with Rubidium. Perhaps even more importantly, the well
established now technique of Feshbach resonance~\cite{FR} (see also
Refs.~\cite{FR2} for work in two-component BECs) can be
used to detune the nonlinear coefficients from this degenerate
case of equal strength and, thus, it is relevant to appreciate
the potential robustness (or lack thereof) of these nonlinear waves
in such settings.

We start by presenting DB solitary waves
%postulating such coherent structures
in explicit
%, generalized
analytical form
and identify the algebraic conditions that need to be satisfied
for the relevant solutions to exist. We solve such algebraic
equations for the characteristic properties of the solutions
and offer an interpretation of the resulting expressions. In addition,
we extract conditions under which such families of solutions will
be possible to sustain. In addition to identifying the relevant
solutions in explicit numerical computations, we are able to
more importantly establish their potential existence/robustness
 in the experimentally relevant
setting of trapped binary condensates. Whenever possible, our
considerations will be fully analytical. Examples of this type will
concern, e.g., the explicit form of the DB solitary waves and their
lattices for general coefficients, or the analysis of the motion of
a single DB for general interactions in the presence of the trap.
However, other aspects of our considerations, such as the stability
of the lattices of such waves in either the homogeneous or the trapped
state will be developed by numerical methods. The combination of
both types of tools will provide us with a broad understanding
of the existence, stability and dynamical properties of the single
DB solitary waves and their multiple generalizations as a function of the
nonlinear inter-atomic interaction strengths.
%In particular, we show
%that numerically exact solutions involving trapped molecules of dark-bright
%solitons (and even quadruplets thereof) can be interpreted as trapped
%versions of the analytically available (lattice)
%profiles obtained herein.

We should note that although in the BEC literature, we are not aware
of any investigations along these analytical lines (the closest
analysis which offers numerical borders of existence of single
dark-bright solitons consists of the work of~\cite{csire}), in the
optics literature, there are some similar studies that we now highlight.
Firstly, it should be noted that these cases do not consider the framework
of a harmonic trap, which is less physically relevant
%as the latter is more artificial
in that context.
A study of
%dark-bright
DB solitary waves for general coefficients has been
conducted in the work of~\cite{ieee}, while periodic solutions, yet
{solely} for the limit of equal nonlinear interactions were
obtained in~\cite{uzunov}.

Our presentation will be structured as follows: in section II, we
will provide the relevant model setup and present the well-known
%dark-bright
DB soliton solutions, as introduced in the Manakov limit
%context by the work of~
(see, e.g., Ref.~\cite{buschanglin}). We will also
explore lattices of such solitary waves in the homogeneous case
near that limit and present
our analytical results for the stability/motion of a single DB solitary wave
in the presence of the trap.
%and for the lattices of such waves, with the adjacent waves
%in the bright component being either in phase or out-of-phase.
In section III, we present our numerical considerations, confirming
the existence of both single and multiple DB solitary wave solutions,
both in the vicinity, as well as far from the Manakov limit, both
in the absence, as well as in the presence of the parabolic trap
confining the atoms.
%and quantifying the characteristics of the obtained
%families without the trap and subsequently connecting the solutions
%to ones obtained in the trapped setting.
Finally, in section IV,
we summarize our findings and propose some challenges for future work.

\section{Model Setup and Analytical Considerations}

We commence our analysis by considering a two-component elongated
(along the $x$-direction) repulsive BEC, composed of two different
hyperfine states of the same alkali isotope. We focus on the experimentally
tractable setting of a highly anisotropic trap, i.e., the longitudinal and transverse
trapping frequencies are such that $\omega_x \ll \omega_{\perp}$.
In this case, the system at hand can be described at the mean-field
level by two coupled Gross-Pitaevskii
equations (GPEs) of the form \cite{emergent}:
\begin{eqnarray}
i\hbar \partial_t \psi_j =
%\nonumber \\
\left( -\frac{\hbar^2}{2m} \partial_{x}^2 \psi_j +V(x) -\mu_j + \sum_{k=1}^2 g_{jk} |\psi_k|^2\right) \psi_j.
\label{model}
\end{eqnarray}
In this model, $\psi_j(x,t)$ ($j=1,2$) denote the mean-field wave functions
of the two components (normalized to the numbers of atoms
$N_j = \int_{-\infty}^{+\infty} |\psi_j|^2 dx$), $m$ is the atomic mass, and
$\mu_j$ are the chemical potentials; furthermore,
$g_{jk}=2\hbar\omega_{\perp} a_{jk}$ are the effective one-dimensional
(1D) coupling constants, with $a_{jk}$ denoting the three $s$-wave scattering
lengths ($a_{12}=a_{21}$) which account for collisions between
atoms belonging to the same ($a_{jj}$) or different ($a_{jk}, j \ne k$) species.
The external trapping potential is parabolic, of the form
$V(x)=(1/2)m\omega_x^2 x^2$. Introducing normalized densities $|u_j|^2=2a|\psi_j|^2$, and
measuring
%the densities ,
length, time and energy in units of
%$2a$,
$a_{\perp} =\sqrt{\hbar/\omega_{\perp}}$,
$\omega_{\perp}^{-1}$ and $\hbar \omega_{\perp}$,
respectively,
% we examine the dimensionless form of
Eq.~(\ref{model}) is expressed in the following dimensionless form:
\begin{eqnarray}
i \partial_t u_1  =& -&\frac{1}{2} \partial_{x}^2u_1  + V(x)u_1
+\left(g_{11}|u_1|^2 + {g}_{12}|u_2|^2 -\mu_{1}\right) u_1,
\label{1}
\\
i \partial_t u_2  =& -&\frac{1}{2} \partial_{x}^2u_2 +V(x)u_2
+ \left({g}_{12}|u_2|^2 + {g}_{22}|u_1|^2- \mu_2 \right) u_2.
\label{2}
\end{eqnarray}
%
%In the above equations, we have used the notation $\Psi_1 = u_1$
%and $\Psi_2 = u_2$. Finally,
The normalized external potential in Eqs.~(\ref{1})-(\ref{2})
assumes the form
\begin{eqnarray}
V(x) = \frac{1}{2}\Omega^{2} x^{2} \label{vd},
\label{vb}
\end{eqnarray}
where $\Omega = \omega_x/\omega_\perp $ represents the normalized trap strength.

\subsection{Single DB soliton in the homogeneous system.}
%without the external potential}

%Our aim in the present work is to
We will now illustrate that solitary waves of
the DB type can in fact be found in an explicit analytical form
even outside of the very special integrable regime of $g_{ij}=1$,
where inverse scattering theory provides such explicit solutions~\cite{APT}.
To that effect, we will consider the analytically tractable case
of $V(x)=0$ (and subsequently illustrate how our results are
modified in the presence of a trap) in Eqs.~(\ref{1})-(\ref{2}),
but maintain as general
coefficients as possible, namely $g_{ij}$ will be arbitrary and
will only be constrained by the conditions for the existence of
our solutions in what follows.%In this case Eqs.~(\ref{1},\ref{2}) take the form:
%More specifically, we examine the dimensionless form of
%Eqs.~(\ref{model}):
%%%%%%%%%%%%%%%%%
%Thus, measuring the densities
%$|\Psi_j|^2$, length, time and energy in units of $2a$, $a_{\perp} =\sqrt{\hbar/\omega_{\perp}}$,
%$\omega_{\perp}^{-1}$ and $\hbar \omega_{\perp}$,
%respectively,  we examine the dimensionless form of
%Eqs.~(\ref{model}):
%%%%%%%%%%%%%%%%%%%%%
%
%\begin{eqnarray}
%i \partial_t u_1  &=& -\frac{1}{2} \partial_{x}^2 u_1
%%\nonumber \\
%+ (g_{11} |u_1|^2 + g_{12} |u_2|^2) u_1,
%\label{deq11}
%\\
%i\partial_t u_2  &=& -\frac{1}{2} \partial_{x}^2u_2
%%\nonumber \\
%+ (g_{12} |u_1|^2 + g_{22} |u_2|^2) u_2,
%\label{deq22}
%\end{eqnarray}
%%

We now seek real, standing-wave solutions of Eqs.~(\ref{1})-(\ref{2}),
with $\partial_t u_j=0$, and
%of the form
%$\partial_t u_j=0$
%$u_1= \exp(-i \mu_1 t) \tilde{u}_1$ and
%$u_2= \exp(-i \mu_2 t) \tilde{u}_2$. With a slight abuse
%of notation, we drop the tilde's (and assume without loss of generality
%that our one-dimensional solutions are real) to
obtain:
\begin{eqnarray}
\mu_1 u_1  &=& -\frac{1}{2}  u_1''
%\nonumber \\
+ (g_{11} u_1^2 + g_{12} u_2^2) u_1,
\label{deq111}
\\
\mu_2 u_2  &=& -\frac{1}{2} u_2''
%\nonumber \\
+ (g_{12} u_1^2 + g_{22} u_2^2) u_2,
\label{deq222}
\end{eqnarray}
where primes denote differentiation with respect to $x$.
We now try explicit analytical solutions in the form of a dark (black) solitary wave
for $u_1$ and a bright solitary wave for $u_2$, namely:
\begin{eqnarray}
u_1 &=& A_1 \tanh(b x),
\label{sDB1} \\
u_2 &=& A_2~ {\rm sech}(b x),
\label{sDB2}
\end{eqnarray}
where $A_1$ and $A_2$ denote the amplitudes of the dark and bright component,
respectively, while $b$ stands for the common inverse width.
Inserting the above expressions into
%We find that these
the equations of motion, we find that the latter are satisfied provided that
%the following
a number of algebraic
conditions hold. More specifically, to satisfy Eq.~(\ref{deq111}), we need:
\begin{eqnarray}
\mu_1 &=& b^2 + g_{12} A_2^2,
\label{deqn1}
\\
b^2 &=& g_{11} A_1^2 - g_{12} A_2^2,
\label{deqn2}
\end{eqnarray}
%
%Similarly,
while to satisfy Eq. (\ref{deq222}), we need to have:
\begin{eqnarray}
\mu_2 &=& -\frac{b^2}{2} + g_{12} A_1^2,
\label{deqn3}
\\
b^2 &=& g_{12} A_1^2 - g_{22} A_2^2.
\label{deqn4}
\end{eqnarray}

We can now suggest a simple way to view the relevant solvability conditions:
one can solve Eqs.~(\ref{deqn1}), (\ref{deqn2}) and (\ref{deqn4}) as
3 linear equations in 3 unknowns ($A_1^2$, $A_2^2$ and $b^2$), provided
that the interactions strengths $g_{ij}$ and the chemical potential $\mu_1$
are set. Then, the remaining Eq.~(\ref{deqn3}) can be used as a closure
condition, self-consistently determining the chemical potential of the
second (bright) component. In this viewpoint, the analytical solution
at hand has the amplitude parameters $A_1$ and $A_2$ determined as:
\begin{eqnarray}
A_1^2 &=& \frac{\mu_1}{g_{11}},
\label{deqn5}
\\
A_2^2 &=& \frac{\mu_1}{g_{11}} \frac{g_{11}-g_{12}}{g_{12}-g_{22}},
\label{deqn6}
\end{eqnarray}
and the inverse width parameter $b$ is determined by:
\begin{eqnarray}
b^2 = \frac{\mu_1}{g_{11}} \frac{g_{11} g_{22} - g_{12}^2}{g_{22} - g_{12}},
\label{deqn7}
\end{eqnarray}
while Eq.~(\ref{deqn3}), with input from (\ref{deqn7}) and (\ref{deqn5})
completes the calculation.

Some important --and physically relevant-- conclusions can be already drawn by this calculation
about the nature of the exact solitary waves obtained
through the above calculation and the constraints on the
existence parameters. In particular, it can be directly seen from
Eq.~(\ref{deqn6}) that the bright component can only survive when
\begin{eqnarray}
{\rm min}(g_{11},g_{22}) < g_{12} < {\rm max}(g_{11},g_{22}).
\label{deqn8}
\end{eqnarray}
Furthermore, it is interesting to also infer from Eq.~(\ref{deqn7})
that if $g_{22}>g_{12}$ (i.e., the second component possesses the largest
scattering length, while the dark soliton is in the first component),
then such exact
DB solitons will {\it only} exist for miscible components, namely
%in the miscible regime,
%side, i.e.
for $g_{11} g_{22} > g_{12}^2$. On the other hand, if $g_{22} < g_{12}$
(i.e., if the first component possesses the largest scattering
length {\it and} is the one holding the dark soliton), then
the above explicit DB solitons
will solely exist for immiscible components, i.e., for
%on the immiscible regime
%side that is for
$g_{11} g_{22} < g_{12}^2$.

\subsection{Lattices of DB solitons}

We now consider two types of lattice generalizations of the relevant
single DB soliton solutions. In the first one, the dark solitons
generalize into the form of a Jacobian elliptic function solution of the
sn-type, while the bright solitons generalize into a cn-type solution.
This suggests that the adjacent solitary waves in this structure
are out-of-phase with respect to each other. In the second generalization,
while the dark solitons preserve the same type of structure, the
bright ones are now of the dn-type, amounting to in-phase bright solitons
in the second component.

\subsubsection{DB soliton lattice with
%{\rm sn}- {\rm cn} Solutions:
out-of-phase bright neighbors}

In this case, for the system of Eqs.~(\ref{deq111})-(\ref{deq222}),
we use the ansatz of the form:
\begin{eqnarray}
u_1 &=& A_1 ~{\rm sn}(b x, k),  \\
u_2 &=& A_2 ~{\rm cn}(b x, k),
\end{eqnarray}
%
%$u_1= A_1 {\rm sn}(b x, k)$, $u_2= A_2 {\rm cn}(b x, k)$.
where $k$ is the elliptic modulus.
In this case, the two resulting algebraic equations stemming from Eq.~(\ref{deq111})
read:
\begin{eqnarray}
\mu_1 &=& \frac{1+k^2}{2} b^2 + g_{12} A_2^2,
\label{deqn9}
\\
k^2 b^2 &=& g_{11} A_1^2 - g_{12} A_2^2.
\label{deqn10}
\end{eqnarray}
Similarly, the conditions stemming from Eq.~(\ref{deq222}) are:
\begin{eqnarray}
\mu_2 &=& \frac{1-2 k^2}{2} b^2 + g_{12} A_1^2,
\label{deqn11}
\\
k^2 b^2 &=& g_{12} A_1^2 - g_{22} A_2^2.
\label{deqn12}
\end{eqnarray}

It is interesting to observe that the special limit case of the hyperbolic
functions, namely $k \rightarrow 1$, naturally asymptotes to the single
DB equations' limit of Eqs.~(\ref{deqn1})-(\ref{deqn4}).
The other relevant limit is the trigonometric one of $k \rightarrow 0$,
which provides sinusoidal and cosinusoidal solutions, respectively, for
the two components;
%, solutions, but
nevertheless, direct inspection of the equations
illustrates that this is so
only at the transition threshold between miscibility and immiscibility
(since it can be directly inferred that such solutions only exist for
$g_{11} g_{22} = g_{12}^2$).

Once again, assuming that
%the perspective of
Eqs.~(\ref{deqn9}), (\ref{deqn10}) and (\ref{deqn12}) constitute
a linear system for $A_1^2$, $A_2^2$
and $b^2$, while Eq.~(\ref{deqn11}) determines $\mu_2$ (for fixed
$\mu_1$ and $g_{ij}$), we find the amplitudes:
\begin{eqnarray}
A_1^2 &=& \frac{2 k^2 (g_{12}-g_{22}) \mu_1}{ (g_{12}^2-g_{11} g_{22})
+ k^2 (2 g_{11} g_{12}-g_{12}^2 - g_{11} g_{22})},
\label{deqn13}
\\
A_2^2 &=& \frac{2 k^2 (g_{11}-g_{12}) \mu_1}{ (g_{12}^2-g_{11} g_{22})
+ k^2 (2 g_{11} g_{12}-g_{12}^2 - g_{11} g_{22})},
\label{deqn14}
\end{eqnarray}
while the (inverse) width parameter $b$ is given by:
\begin{eqnarray}
b^2 = \frac{2 (g_{11} g_{22} - g_{12}^2) \mu_1}{ (g_{12}^2-g_{11} g_{22})
+ k^2 (2 g_{11} g_{12}-g_{12}^2 - g_{11} g_{22})}.
\label{deqn15}
\end{eqnarray}

It is again relevant to attempt to extract the conditions under
which these solutions exist. In particular, the product of
Eqs.~(\ref{deqn13})-(\ref{deqn14}) yields that Eq.~(\ref{deqn8})
is still valid. The product of each of Eqs.~(\ref{deqn13})-(\ref{deqn14})
with Eq.~(\ref{deqn15}) yields once again the conclusion that for
the lattice solutions to exist: if the dark soliton lattice is in the
component with the smaller scattering length, the hyperfine states
need to be miscible (i.e., for $g_{11} < g_{12} < g_{22}$, it must
be $g_{12}^2 < g_{11} g_{22}$). On the other hand, if the dark lattice
is in the component with the larger scattering length, then the
states should be immiscible (i.e., for $g_{22} < g_{12} < g_{11}$, it must
be $g_{12}^2 > g_{11} g_{22}$). Nevertheless, an additional, more
complex condition emerges from the denominator
$D_e=(g_{12}^2-g_{11} g_{22})
+ k^2 (2 g_{11} g_{12}-g_{12}^2 - g_{11} g_{22})$ of the expressions
of Eq.~(\ref{deqn13})-(\ref{deqn15}). In particular,
for $g_{11} < g_{12} < g_{22}$, it must be that $D_e<0$, while
for  $g_{22} < g_{12} < g_{11}$, the opposite must be true, namely
$D_e>0$. By considering this denominator as a binomial in
$g_{12}$, it is clear that $g_{12}$ should be outside the interval
of its roots for $D_e>0$ and inside the same interval for $D_e<0$.
%For the case of parameters such as those of $^{87}$Rb where
%$g_{11}$ and $g_{22}$ are proximal in value, it is straightforward
%to see (by evaluating the roots and analyzing their limits as $k
%\rightarro 0$ and $k \rightarrow 1$) that no additional constraint
%is placed if $g_{11} > g_{22}$ by this condition. On the contrary,
%if $g_{22}>g_{11}$

It is important to note here that no constraint has, a priori, been
placed on the additional parameter , i.e., the elliptic modulus $k$
%, which
appearing in the equations above,
%and is the modulus of the respective elliptic functions,
aside from the requirement that $D_e<0$ or $D_e>0$, depending on the ordering
of the scattering lengths. Nevertheless, $k$ is a critical parameter
since it controls the separation between the solitary waves, which
for the above solution is given by $s=2 K(k)/b$, where $K$ denotes
the complete elliptic integral of the first kind.

\subsubsection{DB soliton lattice with
%{\rm sn}-{\rm dn} Solutions:
in-phase bright neighbours}

We now consider the case where the first component still
has the same profile as in the previous lattice example, namely $u_1=A_1 {\rm sn}(b x, k)$,
while the second component has the form:
\begin{eqnarray}
u_2=A_2 {\rm dn}(b x, k).
\end{eqnarray}
In this case, the solvability conditions
from Eq.~(\ref{deq111}) become
\begin{eqnarray}
\mu_1 &=& \frac{1+k^2}{2} b^2 + g_{12} A_2^2,
\label{deqn16}
\\
k^2 b^2 &=& g_{11} A_1^2 - k^2 g_{12} A_2^2,
\label{deqn17}
\end{eqnarray}
while those stemming from Eq.~(\ref{deq222}) acquire the form:
\begin{eqnarray}
\mu_2 &=& \frac{2-k^2}{2} b^2 + \frac{g_{12}}{k^2} A_1^2,
\label{deqn18}
\\
k^2 b^2 &=& g_{12} A_1^2 - k^2 g_{22} A_2^2.
\label{deqn19}
\end{eqnarray}
Once again the hyperbolic function limit $k \rightarrow 1$ yields
the familiar form of the DB solitary wave solvability conditions.
In this case, the trigonometric limit $k \rightarrow 0$ does not
represent a multi-component solution.

Solving in the familiar way Eqs.~(\ref{deqn16}), (\ref{deqn17}) and
(\ref{deqn19}), we obtain the amplitudes
\begin{eqnarray}
A_1^2 &=& \frac{2 k^2 (g_{12}-g_{22}) \mu_1}{2 g_{11} g_{12} - g_{12}^2 - g_{11}
g_{22} + k^2 (g_{12}^2 - g_{11} g_{22})},
\label{deqn20}
\\
A_2^2 &=& \frac{2 k^2 (g_{11}-g_{12}) \mu_1}{2 g_{11} g_{12} - g_{12}^2 - g_{11}
g_{22} + k^2 (g_{12}^2 - g_{11} g_{22})},
\label{deqn21}
\end{eqnarray}
while the inverse width $b$ parameter is obtained by
\begin{eqnarray}
b^2 = \frac{2 (g_{12}^2 - g_{11} g_{22}) \mu_1}{2 g_{11} g_{12} - g_{12}^2 - g_{11}
g_{22} + k^2 (g_{12}^2 - g_{11} g_{22})}.
\label{deqn22}
\end{eqnarray}

In addition to the constraints of the single DB solitary wave
(obtained as in the previous subsection by pairwise multiplication
of Eqs.~(\ref{deqn20})-(\ref{deqn22})), an additional constraint
stems from the denominator
$\tilde{D}_e= 2 g_{11} g_{12} - g_{12}^2 - g_{11}
g_{22} + k^2 (g_{12}^2 - g_{11} g_{22})$, which should be such that
if $g_{11} < g_{12} < g_{22}$, then $\tilde{D}_e < 0$, while
if $g_{11} > g_{12} > g_{22}$, then $\tilde{D}_e > 0$.
Once again, this can be viewed as a binomial in $g_{12}$ with the
corresponding condition being translated as a statement about the
placement of $g_{12}$ in comparison to its roots.
In this case too, the separation between adjacent
solitary waves is controlled by $k$, with the relevant
distance being $s= 2 K(k)/b$.

\subsection{Dynamics of a single DB soliton in the trap}
%within External Trap: Perturbation Theory}
%We consider a two-component elongated (along the $x$-direction) repulsive BEC, composed of two different
%hyperfine states of the same alkali isotope. Assuming that the trap is highly anisotropic, with the longitudinal and transverse trapping frequencies being such that $\omega_x \ll \omega_{\perp}$, we may describe this system by the following two coupled GPEs:
%%
%\begin{eqnarray}
%i\hbar \partial_t \psi_j =
%\left( -\frac{\hbar^2}{2m} \partial_{x}^2 +V_j(x) -\mu_j + \sum_{k=1}^2 g_{jk} |\psi_k|^2\right)\!\psi_j.
%\label{model}
%\end{eqnarray}
%%
%Here, $\psi_j(x,t)$ ($j=1,2$) denote the mean-field wave functions of the two components (normalized to the
%numbers of atoms $N_j = \int_{-\infty}^{+\infty} |\psi_j|^2 dx$), $m$ is the atomic mass, $\mu_j$ are the chemical potentials, $g_{jk}=2\hbar\omega_{\perp} a_{jk}$ are the effective 1D coupling constants,
%$a_{jk}$ denote the three $s$-wave scattering lengths (note that $a_{12}=a_{21}$) that account for collisions between atoms belonging to the same ($a_{jj}$) or different ($a_{jk}, j \ne k$) species, and $V_j(x)$ represent the external trapping
%potentials.
%
%We assume that both components are confined by the usual harmonic trap, namely
%$V(x)=(1/2)m\omega_x^2 x^2$.
%
% Measuring the densities $|\psi_j|^2$, length, time and energy in
%units of $2a_{11}$, $a_{\perp} = \sqrt{\hbar/\omega_{\perp}}$, $\omega_{\perp}^{-1}$ and $\hbar\omega_{\perp}$,
%respectively,

Finally, from the point of view of analytical considerations,
another case that can be studied
%examined
is that of the dynamics of a single DB soliton
%, but now
in the presence of a parabolic trap. Here, we will resort
to the use of Hamiltonian perturbation theory in order to appreciate
the effect of the trap on the soliton dynamics (see, e.g., \cite{pe2a,vas} and the review \cite{revfr}).
More specifically, we
%may
start by casting Eqs.~(\ref{1})-(\ref{2}) into the following form:
\begin{eqnarray}
i \partial_t u_d  =& -&\frac{1}{2} \partial_{x}^2u_d  + V(x)u_d
%\nonumber \\
%&+&
+ \left(|u_d|^2 + \tilde{g}_{12}|u_b|^2 -\mu_{d}\right) u_d,
\label{deq1}
\\
i \partial_t u_b  =& -&\frac{1}{2} \partial_{x}^2u_b +V(x)u_b
%\nonumber \\
%&+&
+ \left(\tilde{g}_{12}|u_b|^2 + \tilde{g}_{22}|u_d|^2- \mu_b \right) u_b.
\label{deq2}
\end{eqnarray}
In the above equations, we have used the notation $u_1 = u_d$ and $u_2 = u_b$ (and also $\mu_1=\mu_d$ and $\mu_2=\mu_b$), indicating that the component $1$ ($2$) will be supporting a dark (bright) soliton and $\tilde{g}_{12}=\alpha_{12}/\alpha_{11}=\alpha_{21}/\alpha_{11}$,
$\tilde{g}_{22}=\alpha_{22}/\alpha_{11}$.
%
%
%Finally, the external potentials in Eqs.~(\ref{deq1})-(\ref{deq2}) take the form
%%
%\begin{eqnarray}
%V_d(x)&=&V(x) = \frac{1}{2}\Omega^{2} x^{2} \label{vd}, \\
%V_b(x)&=&V(x) = \frac{1}{2}\Omega^{2} x^{2},
%\label{vb}
%\end{eqnarray}
%%
%where $\Omega = \omega_x/\omega_\perp $ represents
% the normalized trap strength.
%
%\subsection{Perturbation theory}
%%%%%%%%%%%%% background  %%%%%%%%%
%
Assuming that the dark soliton is on top of a Thomas-Fermi (TF) cloud characterized by
the density $|u_{\rm TF}|^2=\mu_{d}-V(x)$, we may substitute
%Accordingly,
the density $|u_d|^2$ in
Eqs.~(\ref{deq1})-(\ref{deq2})
%is substituted
by $|u_d|^2 \rightarrow |u_{\rm TF}|^2 |u_d|^2$ \cite{revfr}.
%where $|u_{\rm TF}|^2=\mu_{d}-V$.
Furthermore, introducing the transformations $t \rightarrow \mu_{d} t$,
$x \rightarrow {\sqrt{\mu_{d}}}x$, $|u_b|^2 \rightarrow \mu_{d}^{-1} |u_b|^2$,
we cast  Eqs.~(\ref{deq1})-(\ref{deq2}) into the form:
\begin{eqnarray}
&&i \partial_{t}u_d  +\frac{1}{2} \partial_{x}^2 u_d  -\left(|u_d|^2 +  \tilde{g}_{12}|u_b|^2 -1 \right) u_d = R_d,
\label{eq1d} \\
&&i \partial_{t} u_b  +\frac{1}{2} \partial_{x}^2u_b - \left(\tilde{g}_{12}|u_d|^2 + \tilde{g}_{22}|u_b|^2-\tilde{\mu}\right) u_b = R_b,
\label{eq2d}
\end{eqnarray}
where $\tilde{\mu} =\mu_b/\mu_d $, the functional perturbations $R_d$ and $R_b$ are given by:
\begin{eqnarray}
R_d &\equiv& \left(2\mu_{d}^2\right)^{-1} \Big[2 \left(1-|u_d|^2 \right)V(x)u_d +V'(x)\partial_x u_d \Big],
\label{Rd} \\
R_b &\equiv &\mu_{d}^{-2} \left(1-\tilde{g}_{12}|u_d|^2 \right)V(x)u_b,
 \label{Rb}
\end{eqnarray}
with $V'(x)\equiv dV/dx$.
Equations (\ref{eq1d})-(\ref{eq2d}) can be viewed as a system of two coupled perturbed NLS equations, with perturbations given by Eqs.~(\ref{Rd})-(\ref{Rb}). In the absence of the perturbations
%($\Omega=0$)
%
%and for $\tilde{g}_{ij}=1$, this model constitutes the
%famous integrable two-component (in fact, the generalization can
%take place for N-components) nonlinear Schr{\"o}dinger equation,
%which is often referred to as Manakov model~\cite{manakov}; see also~\cite{APT}.
%We consider the boundary conditions $|u_d|^2 \rightarrow 1 $ and $|u_b|^2 \rightarrow 0$ as $|x| \rightarrow \infty$, under which the NLS Eqs.~(\ref{eq1d})-(\ref{eq2d}) for $\tilde{g}_{ij}=1$ possess an exact
%analytical DB soliton solution of the following form:
%{\bf PGK: I am especially concerned here and throughout this
%calculation for the absence of the nonlinear coefficients
%from the form of the unperturbed DB solution. At most only the
%$\tilde{g}_{12}$ enters the calculation below and I really don't
%see how this is possible (??). Please check !!}
%
it is clear that Eqs.~(\ref{eq1d})-(\ref{eq2d}) possess a stationary single DB soliton
[cf. Eqs.~(\ref{sDB1})-(\ref{sDB2})]. However, as we are interested in studying the dynamics of a moving
single DB soliton in the trap, it is convenient to consider here another, non-stationary DB soliton solution
of Eqs.~(\ref{eq1d})-(\ref{eq2d}), which can be expressed as follows
(see, e.g., Refs.~\cite{buschanglin,vas} for a similar solution, but in the Manakov limit of $g_{ij}=1$):
\begin{eqnarray}
%\!\!\!\!\!\!
u_d(x,t)&=&\cos\phi\tanh \Big[D \Big(x-x_0 (t)\Big) \Big]+i\sin\phi,
\label{dsoliton2}
\\
%\!\!\!\!\!\!
u_b(x,t)&=&\eta\sech \Big[D \Big(x-x_0(t)\Big) \Big]
%\nonumber \\
%&\times&
\times \exp \Big[ikx+i\theta(t)+i \left(\tilde{\mu}-1)t \right) \Big].
\label{bsoliton2}
\end{eqnarray}
Here, $\phi$ is the dark soliton's phase angle, $\cos\phi$ and $\eta$ represent the amplitudes of the dark and
bright solitons, $D$ and $x_0(t)$ denote the inverse width and the center of the DB soliton, while
$k=D\tan\phi = {\rm const}$ and $\theta(t)$ are the wavenumber and phase of the bright soliton,
respectively. Notice that the dark soliton in the above solution may also be a ``gray'' --i.e., a moving--
one (for $0\ne \phi <\pi/2$), which becomes stationary (black) only in the limiting case of $\phi=0$.
In this limit, the solution of Eqs.~(\ref{dsoliton2})-(\ref{bsoliton2}) coincides with the one given
in Eqs.~(\ref{sDB1})-(\ref{sDB2}), with $\mu_1=\mu_d=1$, $A_1=1$, $A_2=\eta$, and $b=D$ (along with the
normalizations of the nonlinearity coefficients described above).

Inserting Eqs.~(\ref{dsoliton2})-(\ref{bsoliton2}) into Eqs.~(\ref{eq1d})-(\ref{eq2d}), we find that the
soliton parameters should satisfy certain conditions --similar to those given in Eqs.~(\ref{deqn1})-(\ref{deqn4}).
In particular,
%The above solution , it can be found that
%%%%%%%%% new %%%%%%%%%
%We find that these equations are satisfied, provided that the following
%conditions hold.
to satisfy Eq. (\ref{eq1d}), we need:
\begin{eqnarray}
%2 &=& D^2 +\cos^2\phi + \tilde{g}_{12} \eta^2
%\label{miu1}
%\\
D^2&=& \cos^2\phi-\tilde{g}_{12}\eta^2 ,
\label{width1}
\\
\dot{x}_0 &=& D\tan\phi,
\label{x0}
\end{eqnarray}
%Similarly,
while to satisfy Eq. (\ref{deq222}), we need to have:
\begin{eqnarray}
%\mu_2 &=& -\frac{D^2}{2}(1-\tan^2\phi) + \tilde{g}_{12}
%\label{miu2}
%\\
D^2&=& \tilde{g}_{12} \cos^2\phi-\tilde{g}_{22}\eta^2,
\label{width2}
\\
\theta(t)&=&\frac{1}{2}\left(D^2-k^2\right)t+\left(1-\tilde{g}_{12}\right)t.
\label{omegat}
\end{eqnarray}
It is clear that the closure conditions of the above equations, namely:
\begin{eqnarray}
%\cos^2\phi  &=& 1
%\label{res1}
%\\
\eta^2 &=& \frac{\tilde{g}_{12}-1}{\tilde{g}_{22}-\tilde{g}_{12}},
\label{res2} \\
D^2 &=& \frac{\tilde{g}_{22}-\tilde{g}_{12}^2}{\tilde{g}_{22}-\tilde{g}_{12}},
\label{res3}
\end{eqnarray}
are consistent with Eqs.~(\ref{deqn6})-(\ref{deqn7}).
%
%We can suggest a simple way to view these solvability conditions.
%One can solve Eqs. (\ref{miu1}), (\ref{width1}) and (\ref{width2}) as
%3 linear equations in 3 unknowns ($\cos^\phi2$, $\eta^2$ and $D^2$), provided
%that the interactions strengths $g_{ij}$ and the chemical potential $\tilde{\mu}_d$
%are set. In this viewpoint, the analytical solution
%at hand has the amplitude parameters $\cos\phi$ and $\eta$ determined as:
%and the inverse width parameter $D$ is determined by:
%\begin{eqnarray}
%
%
%\end{eqnarray}
%
%%%%%%%%%%%%%%%%%%%%%%%%%
%The above parameters of the DB-soliton are connected through the following equations:
%%
%\begin{eqnarray}
%D^2&=& \cos^2\phi-\tilde{g}_{12}\eta^2 = \tilde{g}_{12} \cos^2\phi-\tilde{g}_{22}\eta^2,
%\label{width} \\
%\dot{x}_0 &=& D\tan\phi,
%\label{x0} \\
%\theta(t)&=&\frac{1}{2}\left(D^2-k^2\right)t+\left(1-\tilde{g}_{12}\right)t,
%\label{omegat}
%\end{eqnarray}
%%
%where $\dot{x}_0$ is the DB soliton velocity.
We also note that in our considerations below
we will use the following equation connecting
the number of atoms $N_b$ of the bright soliton with
the amplitude $\eta$ of the bright soliton, the dark-soliton component's chemical potential $\mu_d$, and the
inverse width $D$ of the above DB soliton:
\begin{equation}
N_b\equiv\intl |u_b|^2dx=\frac{2\sqrt{\mu_d}\eta^2}{D}.
\label{Nbmod}
\end{equation}

Let us now assume that the DB soliton evolves adiabatically in the presence of the small perturbation, and employ the Hamiltonian approach of the perturbation theory for matter-wave solitons to study the DB-soliton dynamics. We start by considering the Hamiltonian (total energy) of the system of Eqs.~(\ref{eq1d})-(\ref{eq2d}), when the perturbations are absent ($R_d=R_b=0$), namely,
\begin{eqnarray}
E &=& \frac{1}{2}\int_{-\infty}^{+\infty} \mathcal{E} dx, \nonumber \\
\mathcal{E} &=& |\partial_{x} u_d|^2+|\partial_{x} u_b|^2+\left(|u_d|^2-1\right)^2 %\nonumber \\
%&+&
+ \tilde{g}_{22}|u_b|^{4}-2\tilde{\mu}|u_b|^2+2\tilde{g}_{12}|u_b|^2|u_d|^2.
\label{energy}
\end{eqnarray}
The energy of the system, when calculated for the DB-soliton solution of Eqs.~(\ref{dsoliton2})-(\ref{bsoliton2}), takes the following form:
\begin{eqnarray}
E
%&=&
=\frac{4}{3}D^3+ \frac{1}{6}\chi D^2\left(2\tilde{g}_{12}+3\tan^2\phi +1\right)
%\nonumber \\
%&+&
+ \frac{1}{6}\chi ^2 D\left(\tilde{g}_{22}-\tilde{g}_{12}^2\right)+ \chi\left(\tilde{g}_{12}-\tilde{\mu}\right),
\label{chi}
\end{eqnarray}
where $\chi =N_b/ \sqrt{\mu _d}$.

Since we have considered an adiabatic evolution of the DB soliton, we may assume that, in the presence of the perturbations of Eqs.~(\ref{Rd})-(\ref{Rb}), the DB soliton parameters become slowly-varying unknown functions of time $t$. Thus, the DB soliton parameters become $\phi \rightarrow \phi(t)$, $D \rightarrow D(t)$, and, as a result,
Eqs.~(\ref{width1})-(\ref{x0}) read:
\begin{eqnarray}
D^2(t)&=&\cos^2\phi(t)-\frac{1}{2} \tilde{g}_{12}  \chi D(t),
\label{s1} \\
\dot{x}_0(t)&=&D(t)\tan\phi(t),
\label{s2}
\end{eqnarray}
where we have used Eq.~(\ref{Nbmod}). The evolution system
of the parameters $\phi(t)$, $D(t)$
and $x_0(t)$ can then be closed
by means of the evolution of the DB soliton energy.
In particular, Eq.~(\ref{chi}) with Eqs.~(\ref{s1})-(\ref{s2}) leads to the evolution of the
soliton energy, $dE/dt$. In addition, the latter can be also found using Eqs.~(\ref{eq1d})-(\ref{eq2d})
and their complex conjugates, namely:
%
%it is readily found that
%%
%\begin{eqnarray}
%\frac{dE}{dt} &=& 4D^2\dot{D}+\frac{1}{3}\chi D\dot{D}\left(2\tilde{g}_{12}+3\tan^2\phi +1\right) \nonumber \\
%&+& \chi D^2\tan\phi\sec^2\phi\dot{\phi} + \frac{1}{6} \chi ^2 \dot{D}
%\left(\tilde{g}_{22}-\tilde{g}_{12}^2\right),
%\label{denergy1}
%\end{eqnarray}
%%
%where dots denote derivatives with respect to $t$.
%%
%On the other hand, using Eqs.~(\ref{eq1d})-(\ref{eq2d}) and their complex conjugates, it can be found that
%the evolution of the DB soliton energy, due to the presence of the perturbations, is given by:
%
\begin{eqnarray}
\frac{dE}{dt}&=&-2{\rm Re} \left\{ \int_{-\infty}^{+\infty}
\left( R_d^{\ast}\partial_t u_d + R_b^{\ast}\partial_t u_b \right) dx \right\} \nonumber \\
&=& \frac{V'(x)}{\mu_d^2}\Big[2\sin\phi\cos ^3 \phi -\frac{2}{3}\tilde{g}_{12}\chi D\sin\phi\cos\phi \nonumber \\
&-& \chi D\tan\phi \left(1-\tilde{g}_{12}\left(1-\frac{\cos^2\phi}{3}\right)\right)\Big].
\label{perturb}
\end{eqnarray}
%
%In the case of Eqs~(\ref{Rd})-(\ref{Rb}), the integration of Eq.~(\ref{perturb}) reads:
%%
%\begin{eqnarray}
%\frac{dE}{dt} &=& \frac{V'(x)}{\mu_d^2}\Big[2\sin\phi\cos ^3 \phi -\frac{2}{3}\tilde{g}_{12}\chi D\sin\phi\cos\phi \nonumber \\
%&-& \chi D\tan\phi \left(1-\tilde{g}_{12}\left(1-\frac{\cos^2\phi}{3}\right)\right)\Big].
%\label{de/dt'}
%\end{eqnarray}
%%
%We can then find the evolution of the dark-bright soliton by equating Eq.~(\ref{denergy1}) and~(\ref{perturb}) as follows:
%
Equating the expressions for $dE/dt$, we can end up with the following equation, describing
the evolution of the DB soliton parameters:
\begin{eqnarray}
&4D^2\dot{D}& +\frac{1}{3}\chi D\dot{D}\left(2\tilde{g}_{12}+3\tan^2\phi +1\right)
%\nonumber \\
%&+&
+\chi D^2\tan\phi\sec^2\phi\dot{\phi}
+ \frac{1}{6} \chi ^2 \dot{D}\left(\tilde{g}_{22}-\tilde{g}_{12}^2\right) \nonumber \\
&=& \frac{V'(x)}{\mu_d^2} \Big[2\sin\phi\cos ^3 \phi -\frac{2}{3}\tilde{g}_{12}\chi D\sin\phi\cos\phi
%\nonumber \\
%&-&
- \chi D\tan\phi \left(1-\tilde{g}_{12}\left(1-\frac{\cos^2\phi}{3}\right)\right)\Big].
\label{de/dt}
\end{eqnarray}
The
%last
above equation, together with Eqs.~(\ref{s1})-(\ref{s2}),
%in the form:
%
%\begin{eqnarray}
%&D^2& = \cos^2\phi-\tilde{g}_{12}\frac{N_bD}{2\sqrt{\mu}},
%\label{s11} \\
%&\dot{x}_0(t)& = D(t)\tan\phi(t),
%\label{s22}
%\end{eqnarray}
% 	
form a system of differential equations describing the evolution of
the soliton parameters $\phi$, $D$ and $x_0$. This system can be solved
approximately, upon considering solitons near the center of the trap (i.e., $x_0\approx 0$),
and linearising around the
%the above system has a
fixed point at
\begin{eqnarray}
x_0
%^{(0)}
= 0, \quad \phi_{0}=0, \quad
D_0 = \frac{\chi}{4}\tilde{g}_{12}\left(\sqrt{1+\frac{16}{\chi ^2\tilde{g}^2_{12}}}-1\right).
\label{}
\end{eqnarray}
We can now linearise Eqs.~(\ref{de/dt}) and (\ref{s1})-(\ref{s2}), using the
%following
ansatz: $x_0=
%x_0^{(1)}
X_0$, $\phi=\phi_1$, and $D=D_0+D_1$.
%
%\begin{eqnarray}
%x_0=x_0^{(1)}, \quad \phi=\phi_1, \quad D=D_0+D_1.
%\label{ans}
%\end{eqnarray}
%%
%Inserting the above into Eq.~(\ref{s1}) and neglecting higher order terms we find:
%
%
%We thus obtain the following result:
%\begin{eqnarray}
%D_1&=&-\tilde{D}_0\phi ^2 _1,
%\label{d1}\\
%\tilde{D}_0 &=& \frac{1}{2D_0+\frac{\chi}{2} \tilde{g}_{12}},
%\label{d0tilde}
%\end{eqnarray}
%%	
%
%
%and linearizing Eq.~(\ref{de/dt}) with the use of Eq.~(\ref{d1}) and (\ref{d0tilde}), we obtain by differentiating Eq.~(\ref{x0}) the equation of the motion for the soliton center:
%
To this end, combining the resulting equation for $X_0$, $\phi_1$ and $D_1$, we can end up with the
following equation of motion for the soliton center:
\begin{eqnarray}
\ddot{X}_0&=& -\frac{R}{W}V'(X_0),
\label{exkin}
\end{eqnarray}
where
\begin{eqnarray}
R&=&D_0\left(2-\tilde{g}_{12}\chi D_0+\chi D_0\left(\tilde{g}_{12}-1\right)\right),
\label{R}\\
W &=& 8D^2_0 \tilde{D}_0-\chi D_0^2+\frac{2}{3}\tilde{D}_0 D_0\chi\left(2\tilde{g}_{12}+1\right)
%\nonumber \\
+ \frac{1}{3}\chi^2\tilde{D}_0\left(\tilde{g}^2_{22}-\tilde{g}^2_{12}\right),
\label{W}
\end{eqnarray}
and $\tilde{D}_0 = \frac{1}{2D_0+\frac{\chi}{2} \tilde{g}_{12}}$.
%	
%%
%\begin{figure}
%\begin{tabular}{cc}
%\includegraphics[width=6cm]{g_11_g_22_1_mu_d_15_mu_b_10_omega_01_n_50000.eps}
%\end{tabular}
%\caption{This shows the numerical oscillation frequency through BdG analysis (red solid lines) versus analytical predictions using the Hamiltonian perturbation theory in Eq.~\ref{exkin} (the red solid line is the original version and the green dashed line is the revised one, and the red star representing the prediction from Bush-Anglin for $\Tilde{g}_{11}=\Tilde{g}_{12}=\Tilde{g}_{22}=1$). Here $\Tilde{g}_{22}=1$, $\Omega=0.1$, $\mu_d=1.5$, $\mu_b=1.0$, $n=50000$ and $dx=0.001$. No great difference between the two predictions. And both predictions do not perform well when $\Tilde{g}_{12}$ is far away from 1.}
%\label{frequency_compare}
%\end{figure}
%%
%
%%
%\begin{figure}
%\begin{tabular}{cc}
%\includegraphics[width=6cm]{profiles_stationay_omega_01.eps}
%\end{tabular}
%\caption{This shows the profile of the stationary solution for different $g_{12}$. The parameters are similar with Fig.~\ref{frequency_compare}.}
%\label{stationary}
%\end{figure}
%%
Note that in the Manakov limit
%where
of $\tilde{g}_{12}=\tilde{g}_{22}=1$, Eq.~(\ref{exkin})
%yields to the expected
recovers the equation of motion for the soliton center found in Ref.~\cite{buschanglin}:
\begin{eqnarray}
\ddot{X}_0&=& -\frac{1}{2}V'(X_0) + \frac{N_b}{8\sqrt{\mu + \left(\frac{N_b}{4}\right)^2}}V'(X_0).
\label{reg1}
\end{eqnarray}
%
%\section{Analytical Results}
%
%It is interesting to note that once again, as is well-known also
%for the standard case of $\tilde{g}_{ij}=1$~\cite{buschanglin},
%
In the general case of $g_{ij} \ne 1$, Eq.~(\ref{exkin}) shows that, again, the parabolic trap leads
to a restoring linear force, although here it
%in this case,
is a considerably more complex one, that depends explicitly
on the $\tilde{g}_{ij}$'s. The consequences of this prediction will
be further assessed in the next section, where it will be compared to numerical computations.

\section{Numerical Results}

\subsection{Comparison of Numerics with Analytics}

\subsubsection{Dark-Bright Solitons and Lattices Thereof in
the Homogeneous Case}

 To illustrate the relevance and usefulness
of our analysis, we start the presentation of
our numerical results by a series of computations that
compare the solutions identified numerically with the corresponding
analysis presented above for the homogeneous BEC case,
where the potential
is absent in Eqs.~(\ref{1})-(\ref{2}) i.e., $V(x)=0$. In this
context, we have identified numerically exact solutions (up to a prescribed
precision typically set to $10^{-7}$), using a fixed point iteration
scheme of the Newton-Raphson type. In so doing, we have confirmed
that our analytical solutions are indeed numerically
exact, up to the local truncation
error (of O$(\Delta x^2)$, where $\Delta x$ is the spatial grid discretization
step that enters the numerical computation).

\begin{figure}
\begin{tabular}{cc}
 \includegraphics[width=8cm]{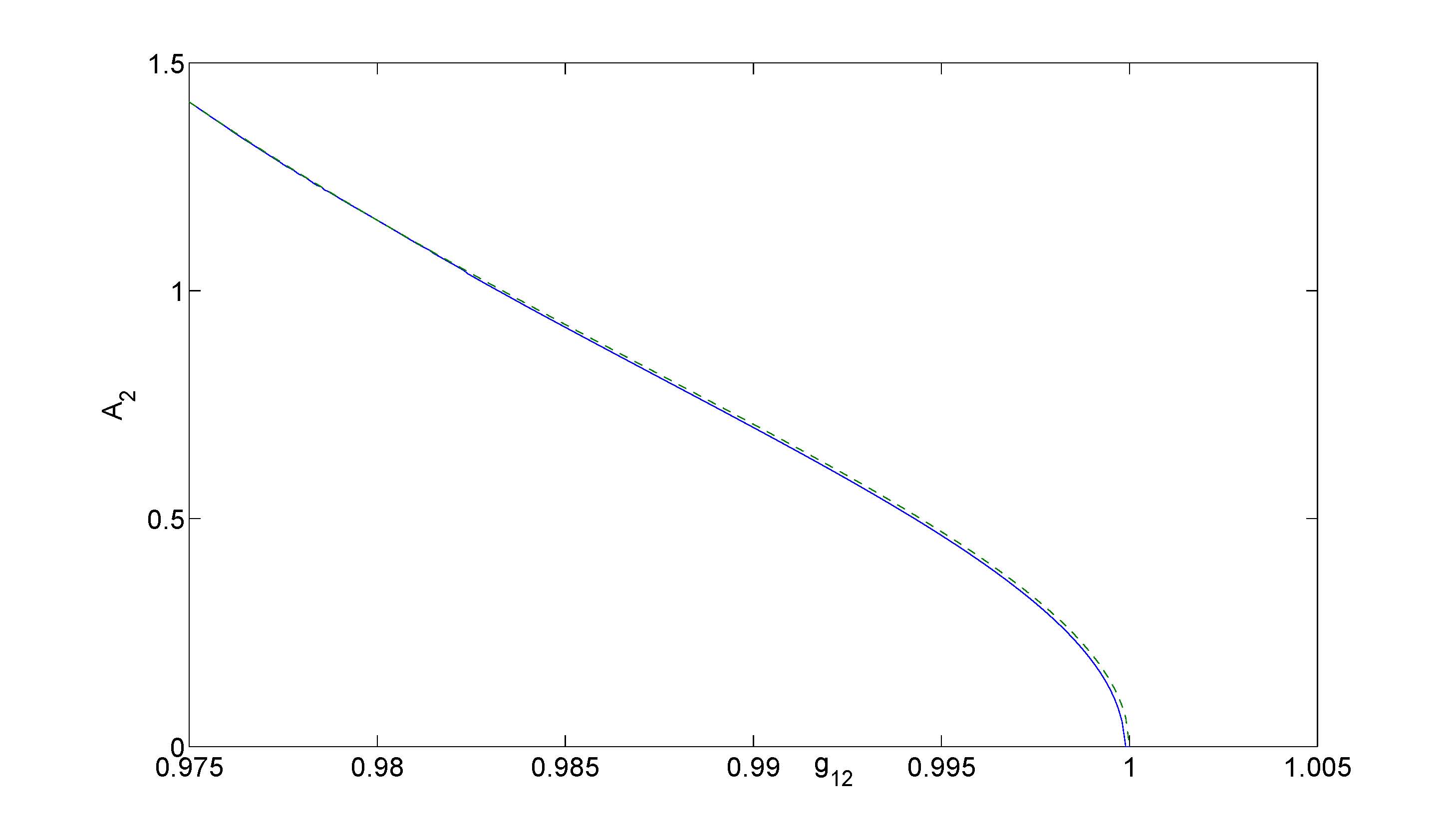} &
 \includegraphics[width=8cm]{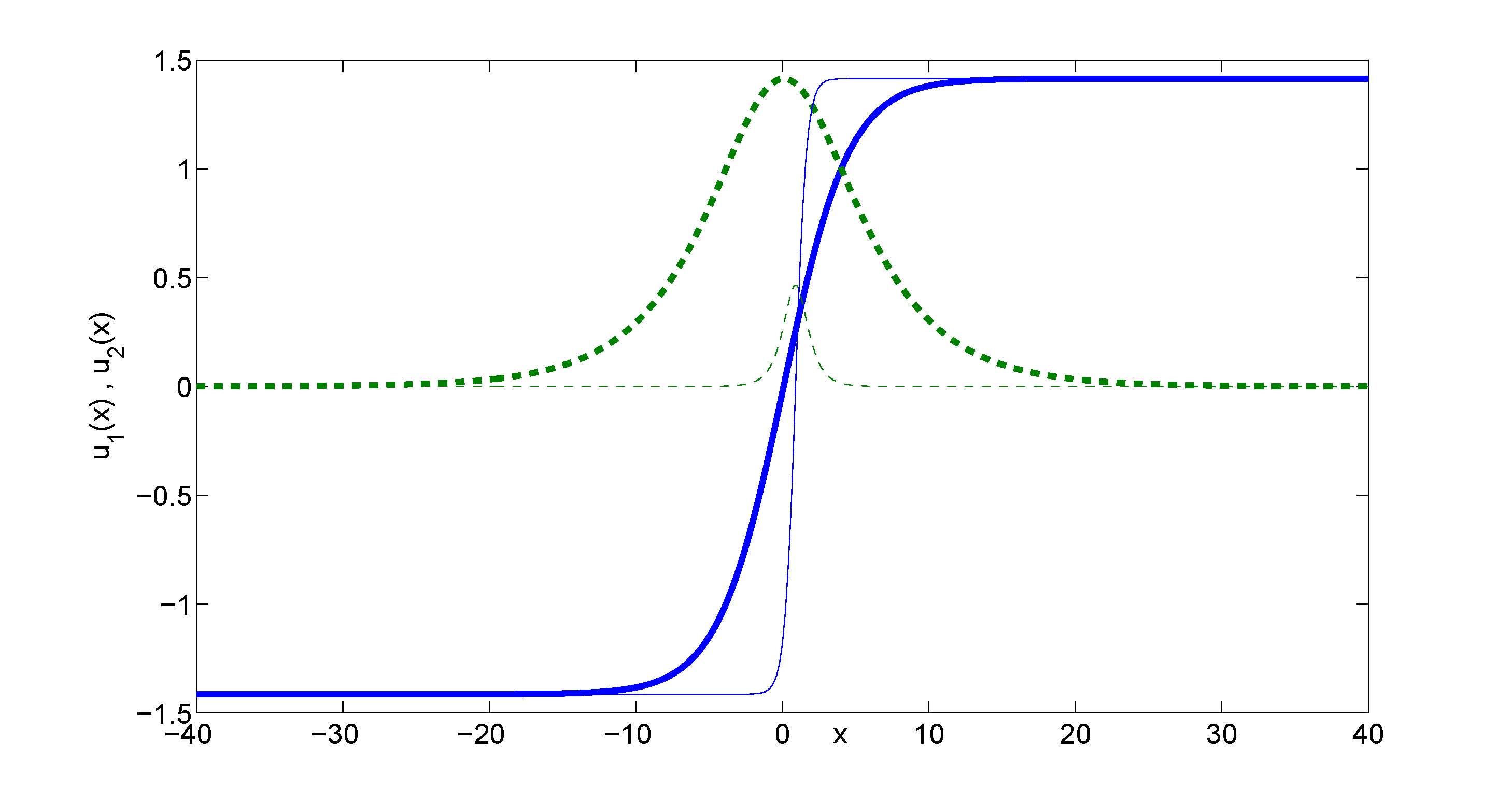} \\
\end{tabular}
\caption{A prototypical example of the comparison of the solution
obtained analytically as a function of continuation in $g_{12}$ for
fixed $g_{11}=1$ and $g_{22}=0.95$, starting with the relevant parameters
for $^{87}$Rb of $g_{12}=0.975$ and approaching
the limit of $g_{12} \rightarrow g_{11}$. The comparison made here concerns
the amplitude $A_2$ of the bright soliton.
% (which we have identified as
%the simplest non-trivial diagnostic to comare; for $A_1$ the
%agreement is naturally excellent but trivial as there is no functional
%dependence).
The dashed line contains the analytical prediction of
Eq.~(\ref{deqn6}), while the solid line is the fully numerical result
obtained as a result of a fixed point iteration in a grid of spacing
$\Delta x=0.2$. The very slight (nearly imperceptible)
disparity stems from local truncation
error (of O$(\Delta x^2)$) of the numerical method. The right panel
contains the numerically obtained (but matching the analytical up
to the local truncation error) dark-bright soliton for $g_{12}=0.975$
(thicker lines; solid for the dark and dashed for the bright)
and for $g_{12}=0.995$ (thinner lines).}
\label{fig_1db}
\end{figure}

This is shown for the case of the
%dark-bright
DB solitary wave
in Fig.~\ref{fig_1db}, where we have fixed the parameters
$g_{11}=1$ (this means that $g_{ij}=\tilde{g}_{ij}$) and $g_{22}=0.95$ to the ones relevant for
$^{87}$Rb;
%while
furthermore, the coefficient $g_{12}$ is initialized weakly on the
immiscibile side at $g_{12}=0.975$ (as is relevant for this
atomic gas), and the variation of the relevant solution is
followed over the range of parameters $g_{12} \in [0.975,1]$.
To confirm that as the inter-species interaction is varied
the analytical solution is followed, we have used --as the simplest
non-trivial diagnostic--
%namely
the amplitude of the
bright component $A_2$ (for $A_1$ the
agreement is naturally excellent, but trivial, as there is no functional
dependence). This is shown in the left panel of the figure, with
the numerical results given by the solid line, while the analytical
expression of Eq.~(\ref{deqn6}) is shown by the dashed one.
On the other hand, the right panel illustrates the nature of the
variation of the solution as the limit of vanishing amplitude
is approached;
in this case, this limit is $g_{12}=g_{11}$, since $g_{11}>g_{22}$ and the dark
soliton is in the component with the largest scattering length.
For increasing $g_{12}$ approaching $g_{11}$, the width of the dark
soliton decreases and, together with it,
%decreases also
the width of the
``trapped'' bright soliton bound state also decreases. In addition, the amplitude of
the bright soliton (proportional to
%the
$\sqrt{g_{11}-g_{12}}$ according
to Eq.~(\ref{deqn6})) also decreases and tends to $0$ at the
relevant limit.

\begin{figure}
\begin{tabular}{cc}
 \includegraphics[width=8cm]{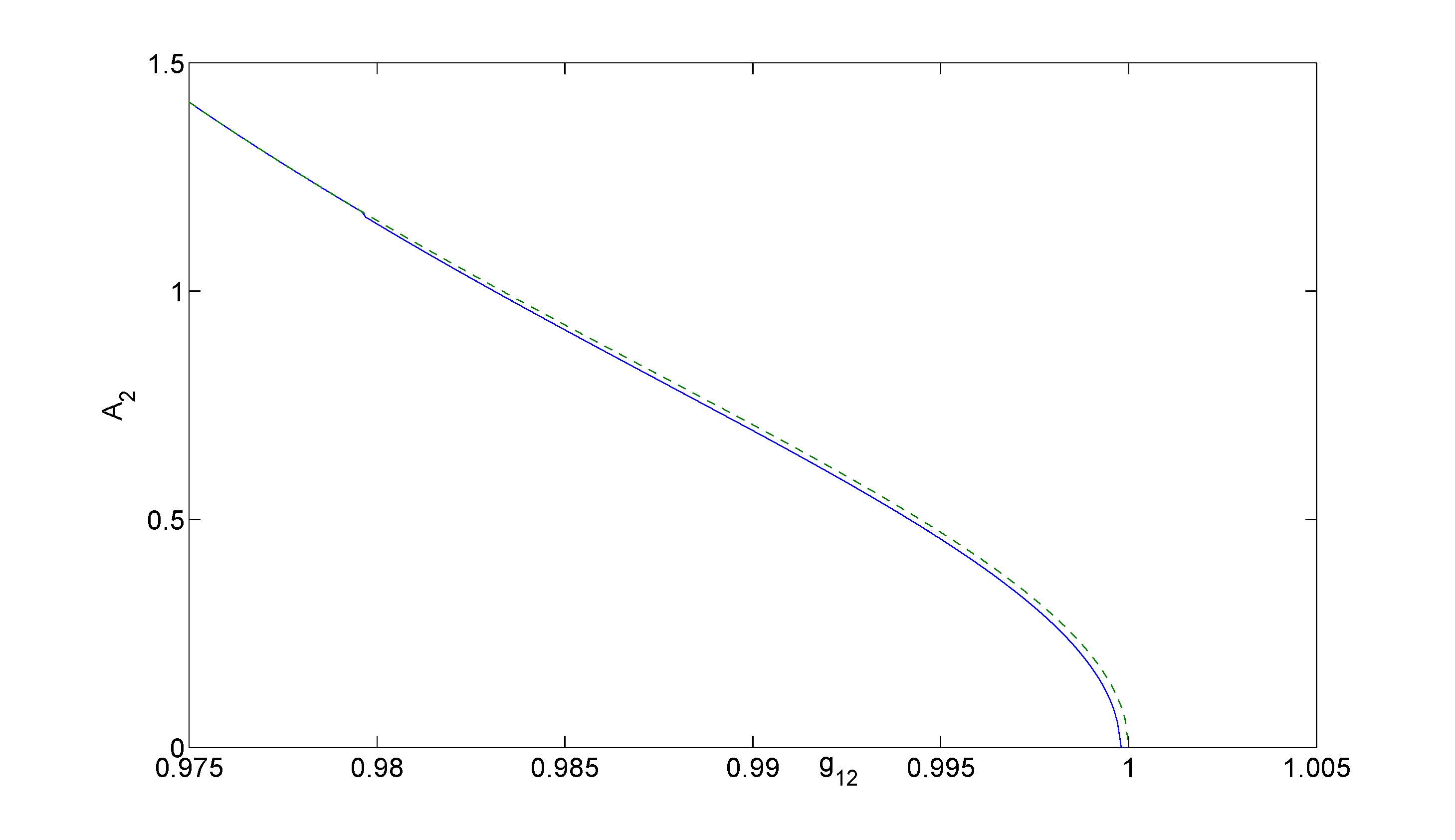} &
 \includegraphics[width=8cm]{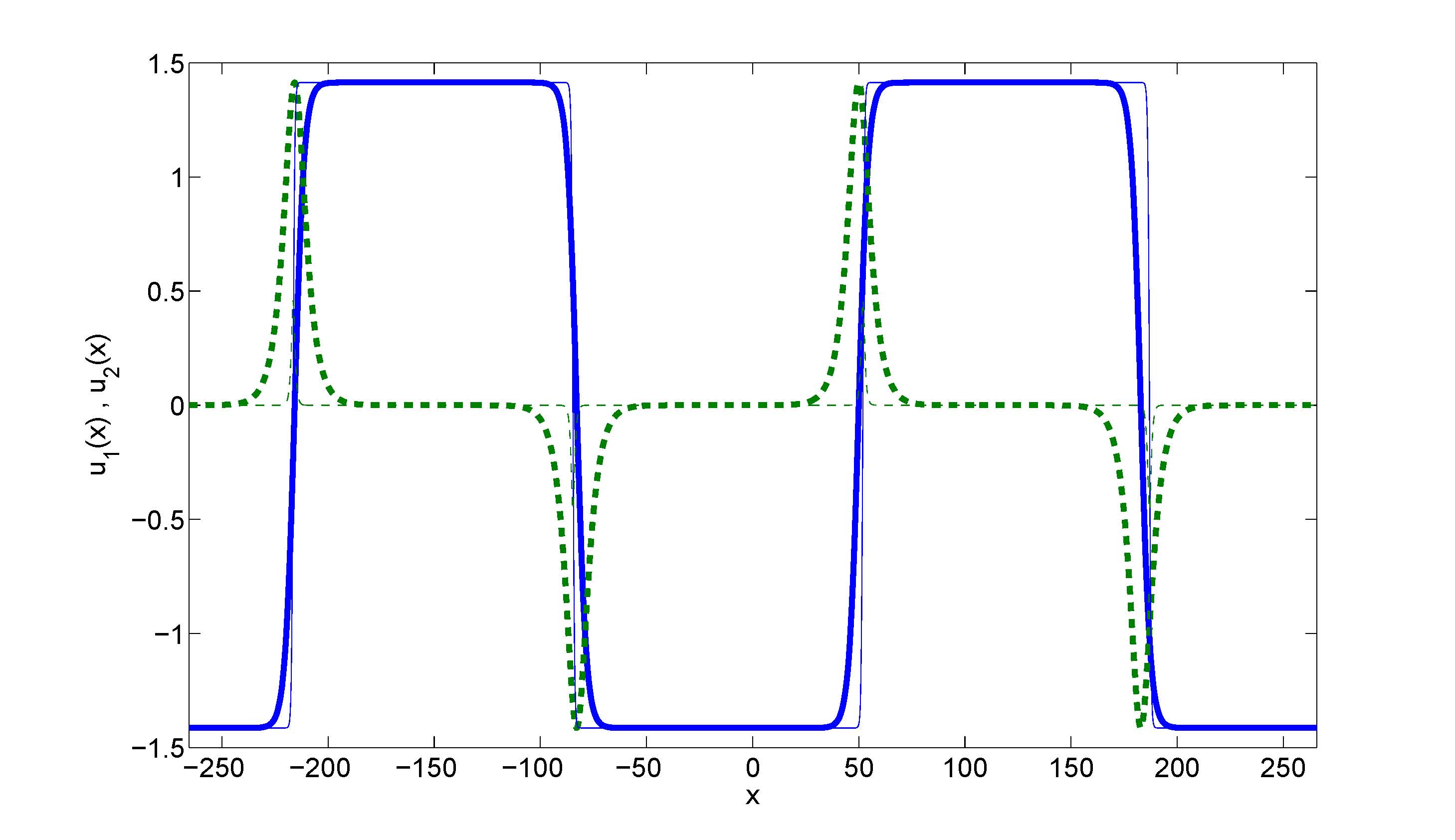} \\
\end{tabular}
\caption{The same diagnostics as for the single dark-bright soliton
of
Fig.~\ref{fig_1db} are used but now for the case of the sn-cn solution branch.}
\label{fig_4db_op}
\end{figure}

\begin{figure}
\begin{tabular}{cc}
 \includegraphics[width=8cm]{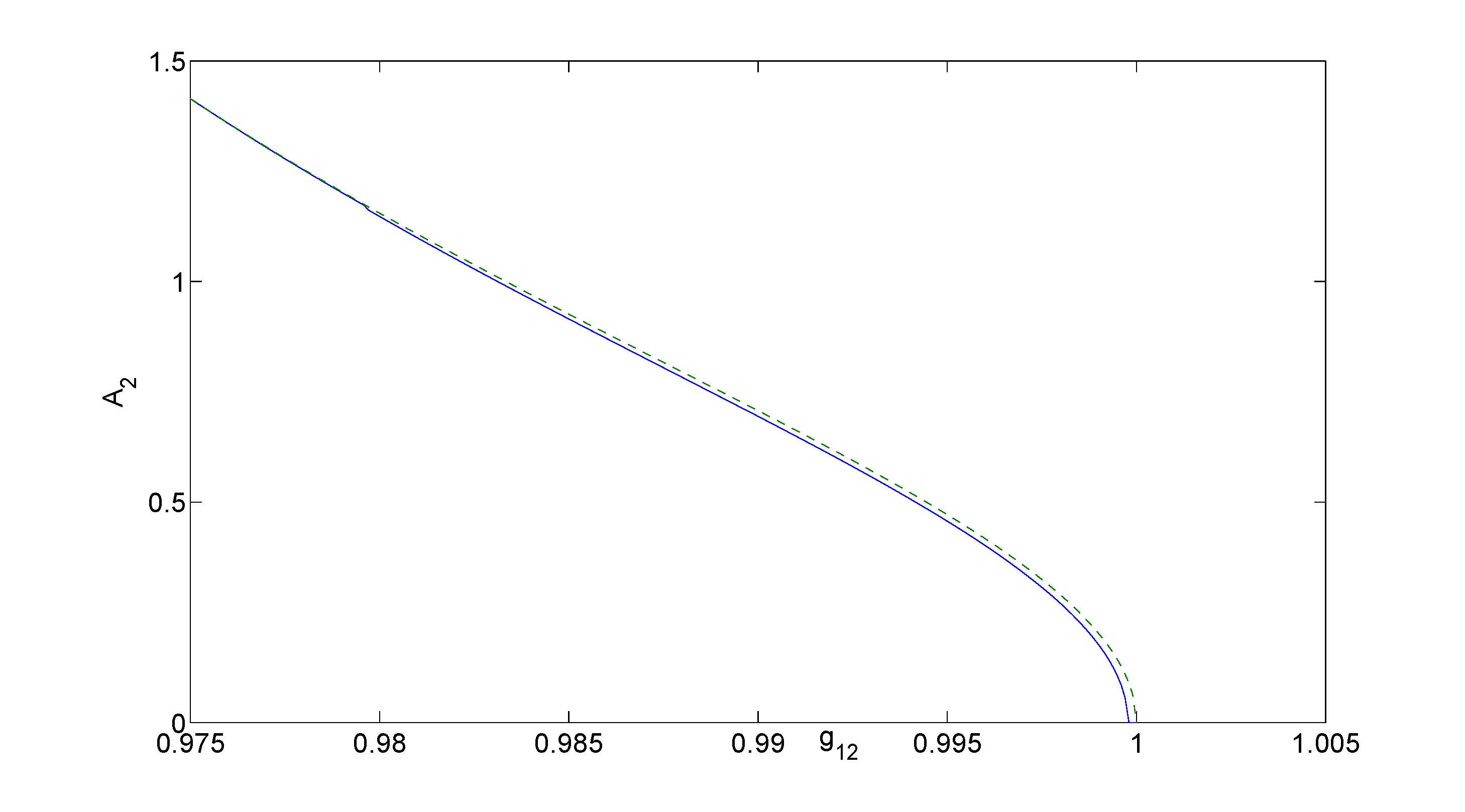} &
 \includegraphics[width=8cm]{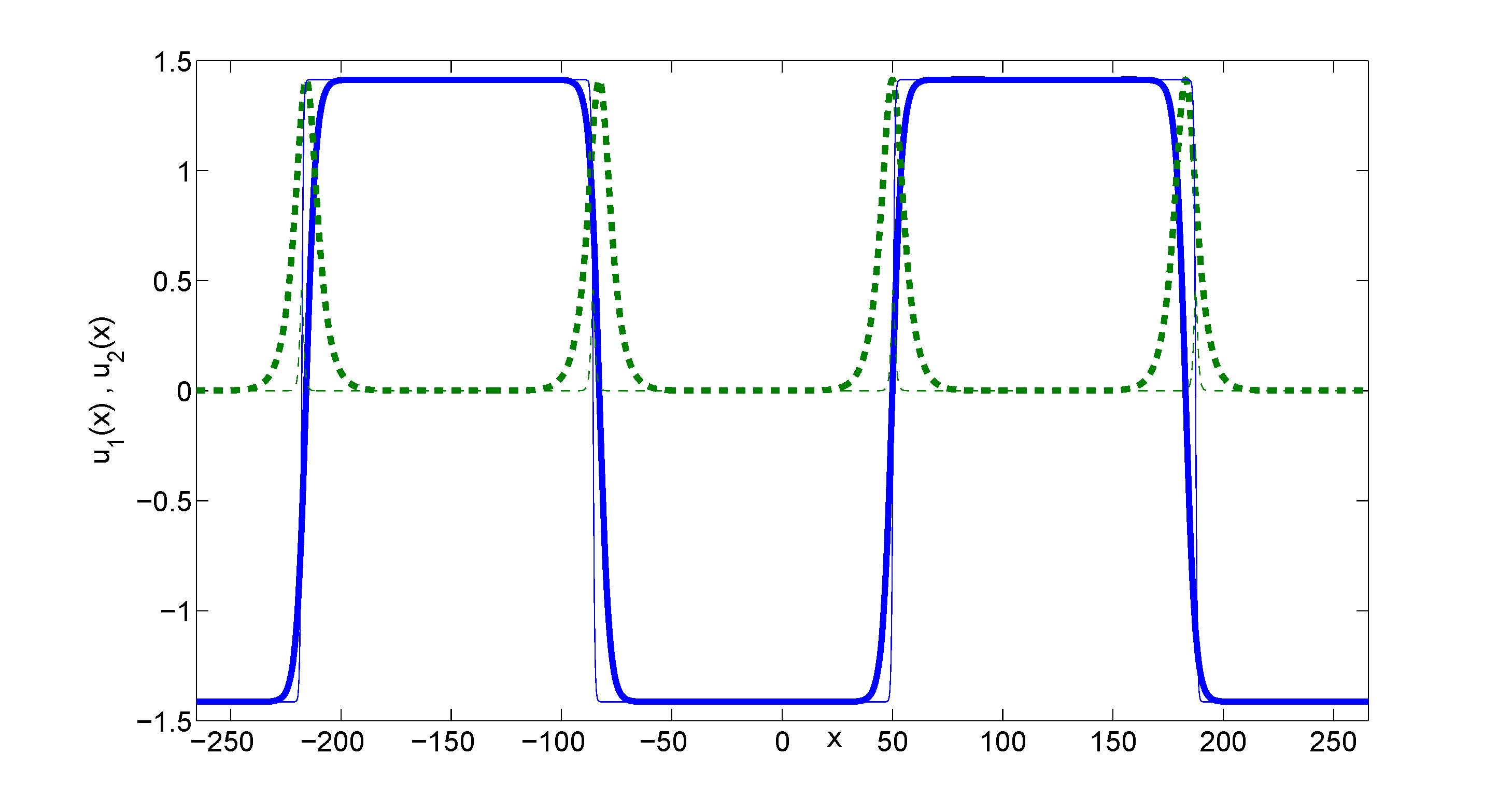} \\
\end{tabular}
\caption{The same diagnostics as for the single dark-bright soliton
of Fig.~\ref{fig_1db} are used but now for the case of the sn-dn solution branch.}
\label{fig_4db_ip}
\end{figure}

Similar diagnostics but now in the case of the soliton lattices
are shown in Figs.~\ref{fig_4db_op}-\ref{fig_4db_ip}. The
former presents the sn-cn solutions, where the bright lattice
bears out-of-phase nearest neighbors, while the latter concerns
the sn-dn case with the bright solitons being all in phase.

\subsubsection{Single DB Soliton in the Presence of a Trap}

\begin{figure}
\begin{tabular}{cc}
\includegraphics[width=8cm]{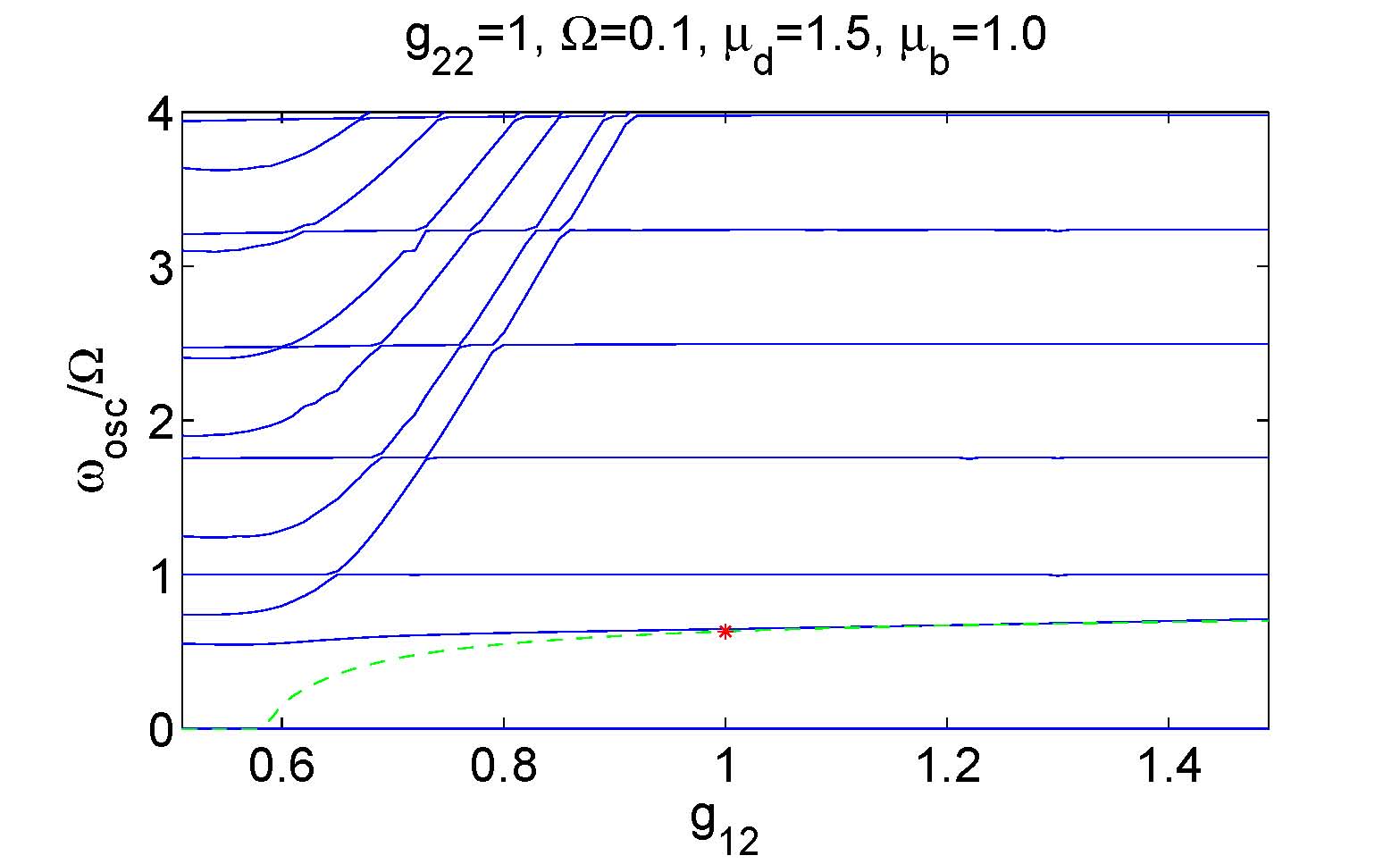}
\end{tabular}
\caption{The figure shows the numerical oscillation frequency through BdG analysis (blue solid lines) versus the analytical predictions using the Hamiltonian perturbation theory in Eq.~(\ref{exkin})
%(the red solid line is the original version and the
(green dashed line),
%is the revised one,
while the red star
represents the prediction from~\cite{buschanglin}
for ${g}_{11}={g}_{12}={g}_{22}=1$). Here ${g}_{22}=1$, $\Omega=0.1$, $\mu_d=1.5$, $\mu_b=1.0$, and $dx=0.001$. Notice that the spectrum
in addition to this anomalous mode of oscillation, bears a large number of
modes (nearly flat) associated with the dark component and a similarly
large number of modes associated with the bright component (bearing a
rapid variation). The theoretical prediction for the anomalous mode is
very good roughly for ${g}_{12} \in [0.8,2]$, while it becomes
progressively worse for lower parameter values.}
%No great difference between the two predictions. And both predictions do not perform well when $\Tilde{g}_{12}$ is far away from 1.}
\label{frequency_compare}
\end{figure}

Our other analytical prediction concerns Eq.~(\ref{exkin})
providing a prediction for the frequency of oscillation of
a DB soliton in the presence of a parabolic (magnetically induced)
trap. While the equation more generally connects the DB motion
through an effective mass to the gradient of the trapping
potential, in the present setting  we will restrict our
considerations to the linear restoring force in the case
of a harmonic trap. To examine the validity of this prediction,
we find the numerically exact (up to the prescribed accuracy
discussed above) solitary wave for different values of
$g_{12}$ (we now fix $\mu_d$ and $\mu_b$, while varying $g_{12}$)
and compare the spectrum of the linearization around
it with the frequency predicted by Eq.~(\ref{exkin}). As argued
in our earlier work (see e.g.~\cite{pe2a}, for
${g}_{ij}=1$), the spectrum of the linearization around
a DB solitary wave should contain an anomalous/negative energy
mode with a frequency associated with the oscillational
frequency of the DB within the parabolic trap. Indeed,
as is confirmed by Fig.~\ref{frequency_compare},
such a frequency is present in this case as well and is found
to be in very good agreement with our theoretical prediction
for this motion in the interval $g_{12} \in [0,2]$. However,
for lower values of the parameter, a progressive discrepancy
between the theoretical prediction and the numerical result
can be discerned e.g. for $g_{12} < 0.8$.

\begin{figure}
\begin{tabular}{cc}
\includegraphics[width=8cm]{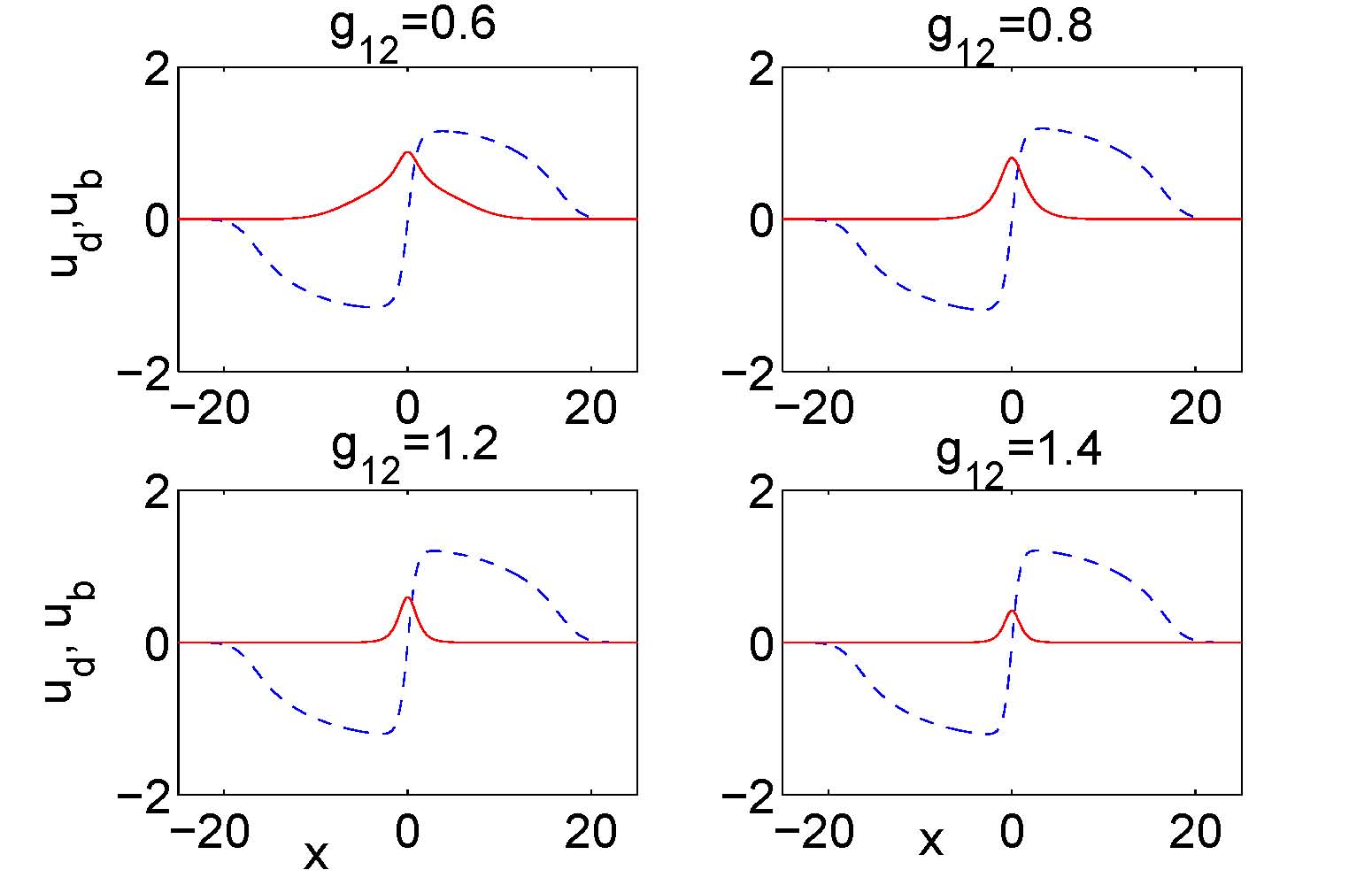}
\end{tabular}
\caption{This shows the profile of the (single wave)
stationary solution for different $g_{12}$.
The parameters are similar with Fig.~\ref{frequency_compare}.}
\label{stationary}
\end{figure}

In an attempt to appreciate the origin of this discrepancy, we illustrate
the form of the solution as $g_{12}$ is decreased in Fig.~\ref{stationary}.
From these findings, it is immediately evident that while our DB ansatz
correctly captures the relevant waveform near and beyond the threshold
for immiscibility, yet, it is far less adequate in describing the solitary
wave on the miscible side. There, the miscible interaction with the
dark component rapidly widens the bright counterpart (see especially
the top left panel of the figure for $g_{12}=0.6$), clearly
illustrating the inadequacy of our hyperbolic secant waveform.
This naturally justifies the interval of good agreement between
the theoretical and numerical oscillation frequency result.

\subsection{Further Numerical Findings}

We now explore more broadly the nature of the solitary DB waves
and of the lattices thereof both in the absence and in the
presence of the trap for features/regimes which are not
captured by our analytical considerations.

\begin{figure}
\begin{tabular}{cc}
\includegraphics[width=8cm]{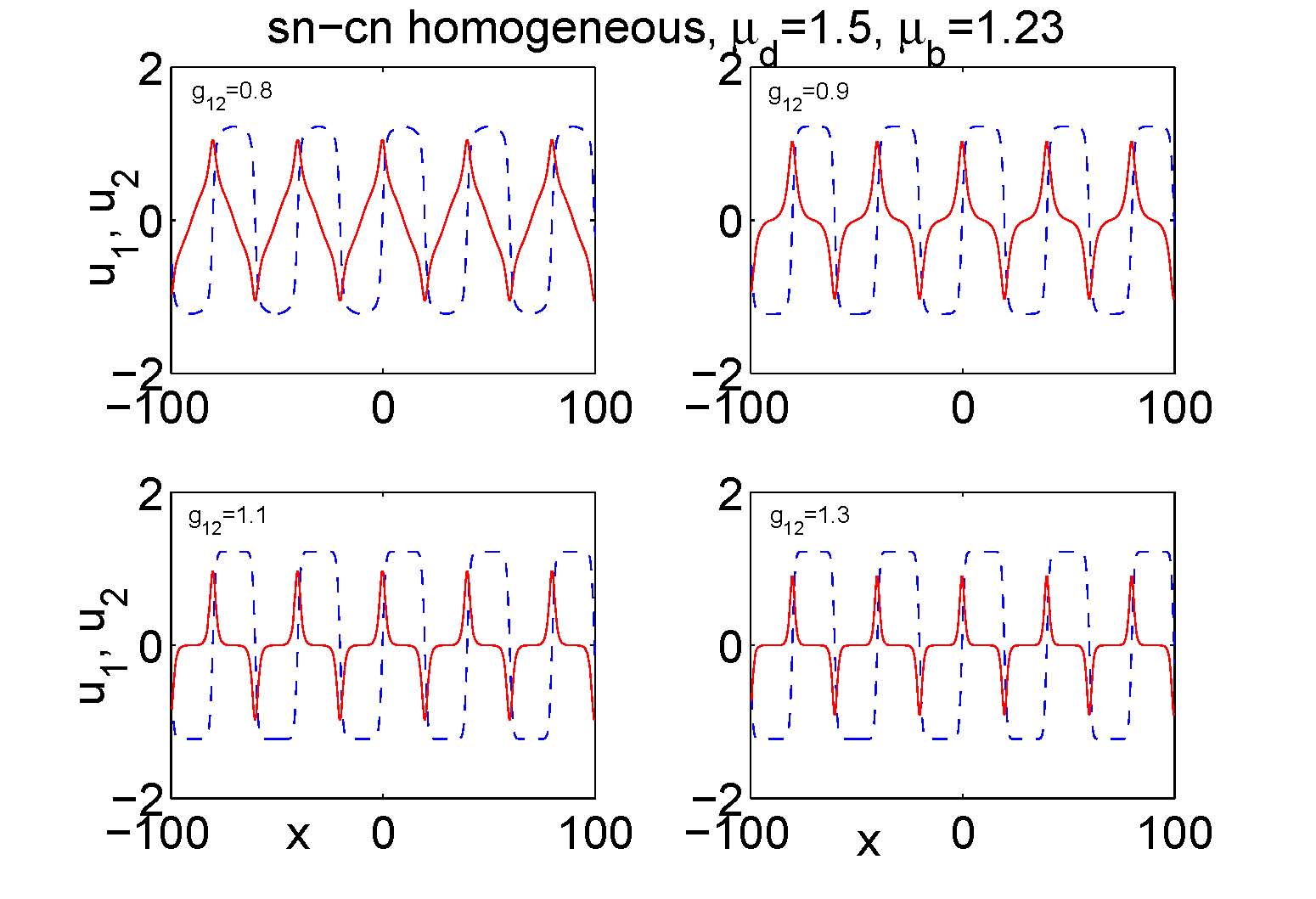}
\end{tabular}
\caption{The figure shows the stationary profile of sn-cn type
periodic solution for $g_{12}=0.8, 0.9, 1.1, 1.3$
on the top left, top right, bottom left and bottom right
panel respectively. The chemical potentials used are $\mu_d=1.5$
and $\mu_b=1.23$.}
%As we can see, when $g_{12}$ decreases to $0.8$ or so, the solution is more like a dark-
%anti-dark pair.}
\label{sn_cn_homogeneous_stationary}
\end{figure}

In Fig.~\ref{sn_cn_homogeneous_stationary}, we now fix the values of
the chemical potentials (at $\mu_d=1.5$ and $\mu_b=1.23$, and $g_{22}$ will be set to 0.95 for computations hereafter)
%{\bf PGK:
%what is the value of ${g}_{22}$ ???})
and vary the value of
${g}_{12}$ from $0.8$ (top left) to $0.9$ (top right), to
$1.1$ (bottom left) to $1.3$ (bottom right). We can see that even
in this region of ${g}_{12}$ which is outside the range of
our analytically tractable lattice solutions of the sn-cn type,
such solutions can still be retrieved numerically. In the immiscible
regime, the solutions consist of thin DB solitons, wherein the
bright components of the pair alternate in phase. The immiscibility
leads the bright component to lie very close to $0$ density in between
its spikes due to the strong mutual repulsion with the finite
density (in these intermediate regions) dark component. However, as
the miscible limit is approached and eventually traversed, while
the dark component does not change significantly, the bright
component broadens considerably and starts approaching a more
``trigonometric'' rather than ``hyperbolic secant'' type shape
between its local maxima/minima.

% Stationary solution: sn-cn type homogeneous
%
% Spectrum: sn-cn type homogeneous FDM
%
\begin{figure}
\begin{tabular}{cc}
\includegraphics[width=9cm]{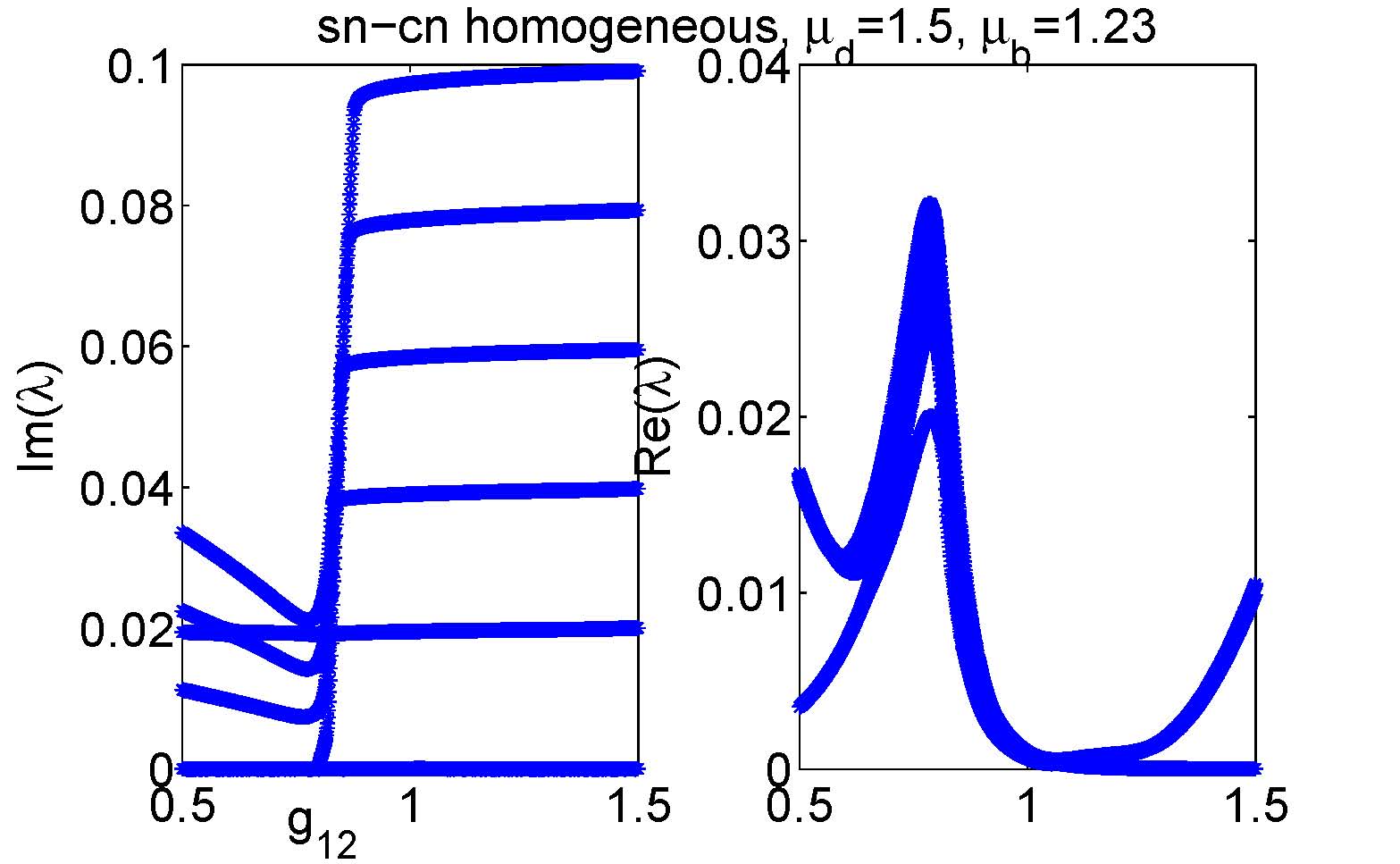}  
\includegraphics[width=8cm]{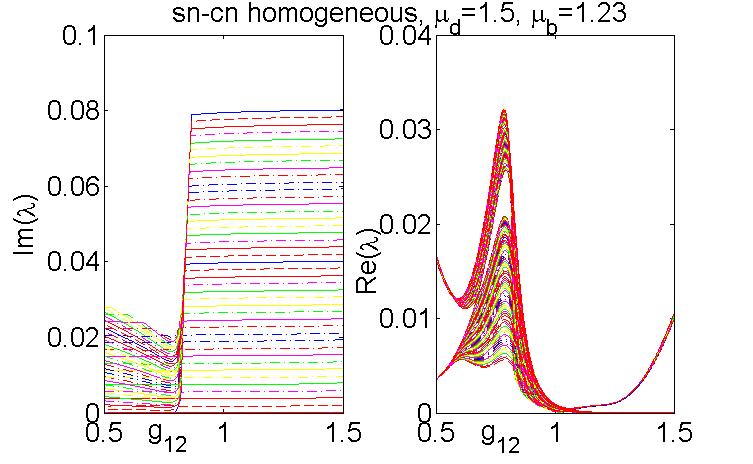}  
\end{tabular}
\caption{The left panel shows the spectrum of the {\rm sn}-{\rm cn} periodic solution as a function of ${g}_{12}$
for $\mu_d=1.5$ and $\mu_b=1.23$ using a finite difference method.
The right panel shows the same spectrum, but also when applying
the so-called Hill's method (using different wavenumbers through imposing
a suitable phase $\theta$ at the edge of a single period and considering
--in this case 11-- different values of $\theta$).}
\label{sn_cn_homogeneous_spectrum_g12_vary}
\end{figure}

We subsequently also examined the linearization spectrum (so-called
Bogolyubov-de Gennes or BdG) spectrum around such a periodic solution,
in order to identify the stability of these states. The
conclusions of our analysis are shown in
Fig.~\ref{sn_cn_homogeneous_spectrum_g12_vary}. The spectrum is obtained
with two methods. The first one, shown in the left panel, concerns the
direct eigenvalue computation of the linearization matrix
around the exact periodic solution that
is obtained from our Newton-Raphson method (with finite differences
applied for the spatial discretization). The second plot of the right
panel ``enhances'' this spectrum by considering the so-called Hill's
method~\cite{hills}, taking direct advantage of the fact that the
solution is periodic to resolve more adequately the perturbation
wavenumbers
associated with the unit cell of its periodicity. This enhancement of
the finite difference method by its combination with the Hill method
has been described in~\cite{hills} and is directly applied here.
We can see that the spectrum derived as a result contains as a part
the linearization spectrum of the left panel, but also fills in
additional eigenvalues due to its ability to more finely probe the
perturbation wavenumbers in comparison to the standard finite
difference scheme.
%This is the result that we get when we use the Hill's
%analysis on those states.
The details of Hill's method are
described in the appendix.

The relevant conclusions are also interesting from a physical point of
view. It can already be seen from the imaginary
parts of the relevant eigenfrequencies that there is a drastic
change of the eigenvalue behavior and of their relative frequency
spacing as the miscible threshold is approached. However, more
critically for our stability purposes, we can observe that
there is an interval of ${g}_{12}$'s in the vicinity
of the miscibility-immiscibility threshold, and especially so
weakly on the immiscible side (i.e., for $1 < {g}_{12} < 1.2$
or so), where the relevant periodic solution is {\it least unstable}.
We should remind the reader that in the Hamiltonian system
considered herein,
instability (at the linearization level) arises whenever an eigenmode
exists with Re$(\lambda) =\neq 0$.
Hence, the potential manipulation of the relevant inter-species
interaction coefficient would be most likely to produce such
long lived solutions on the weakly immiscible side.

%
% sn-dn homogeneous
%
\begin{figure}
\begin{tabular}{cc}
\includegraphics[width=8cm]{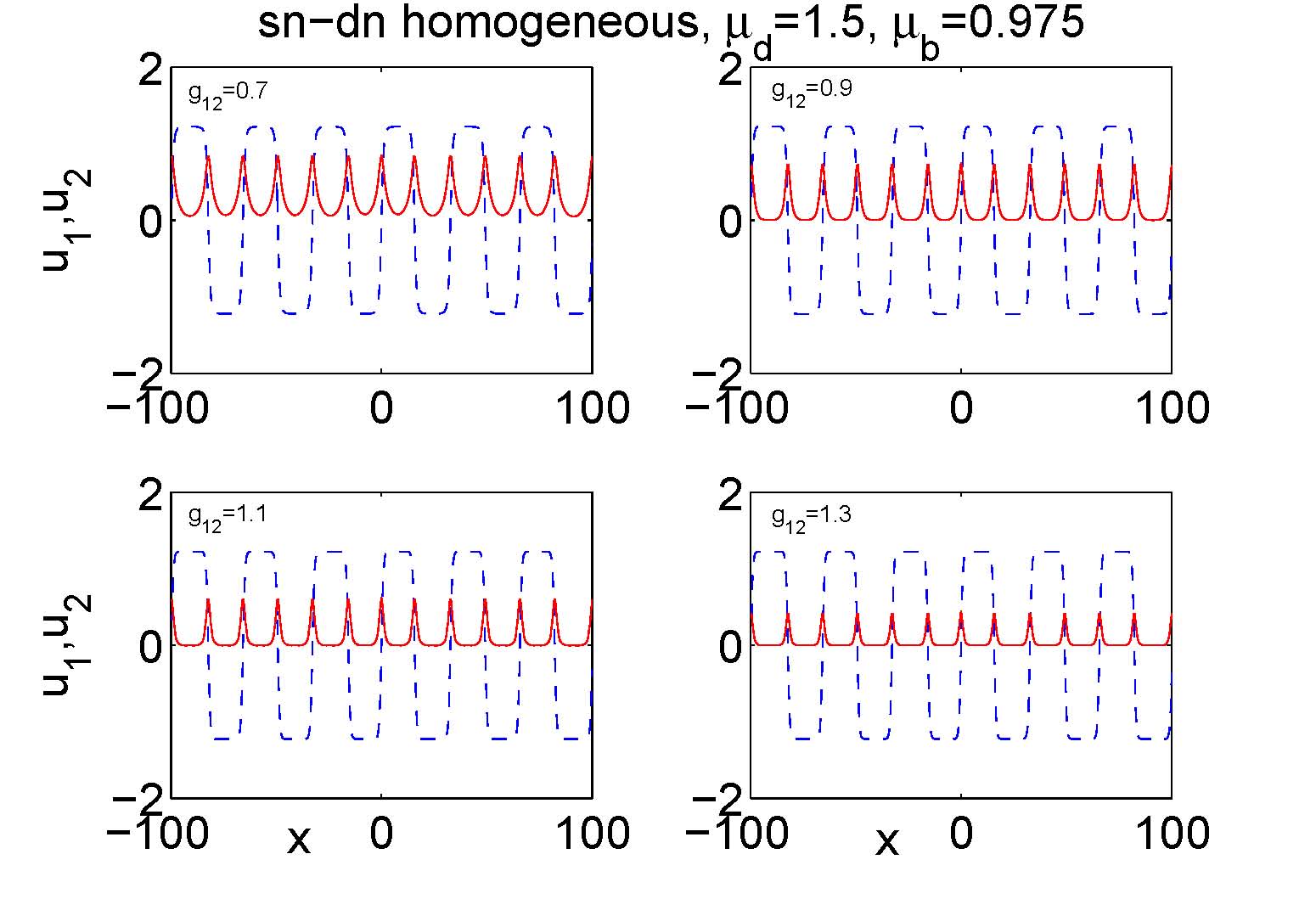}
\end{tabular}
\caption{This shows the stationary profile of sn-dn type periodic solution for $g_{12}=0.7, 0.9, 1.1, 1.3$
on the top left, top right, bottom left and bottom right panel respectively. The chemical potentials
are $\mu_d=1.5$ and $\mu_b=0.975$. For $g_{12}<0.7$ when it is small enough, we see the dn solutions will no
longer touch the x-axis, but rather ``lift up'' above it.}
\label{sn_dn_homogeneous_stationary}
\end{figure}

\begin{figure}
\begin{tabular}{cc}
\includegraphics[width=8cm]{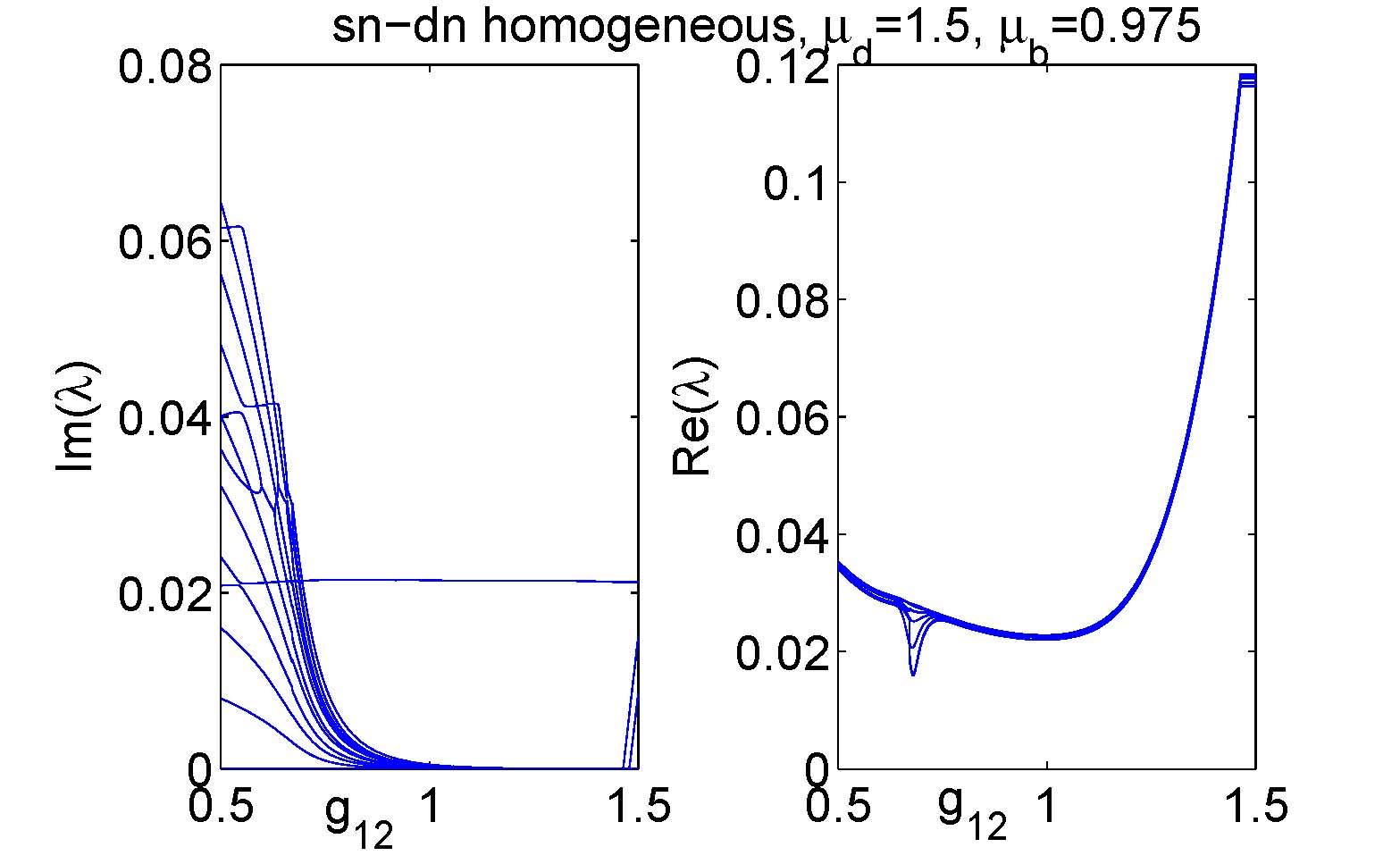}
\includegraphics[width=8cm]{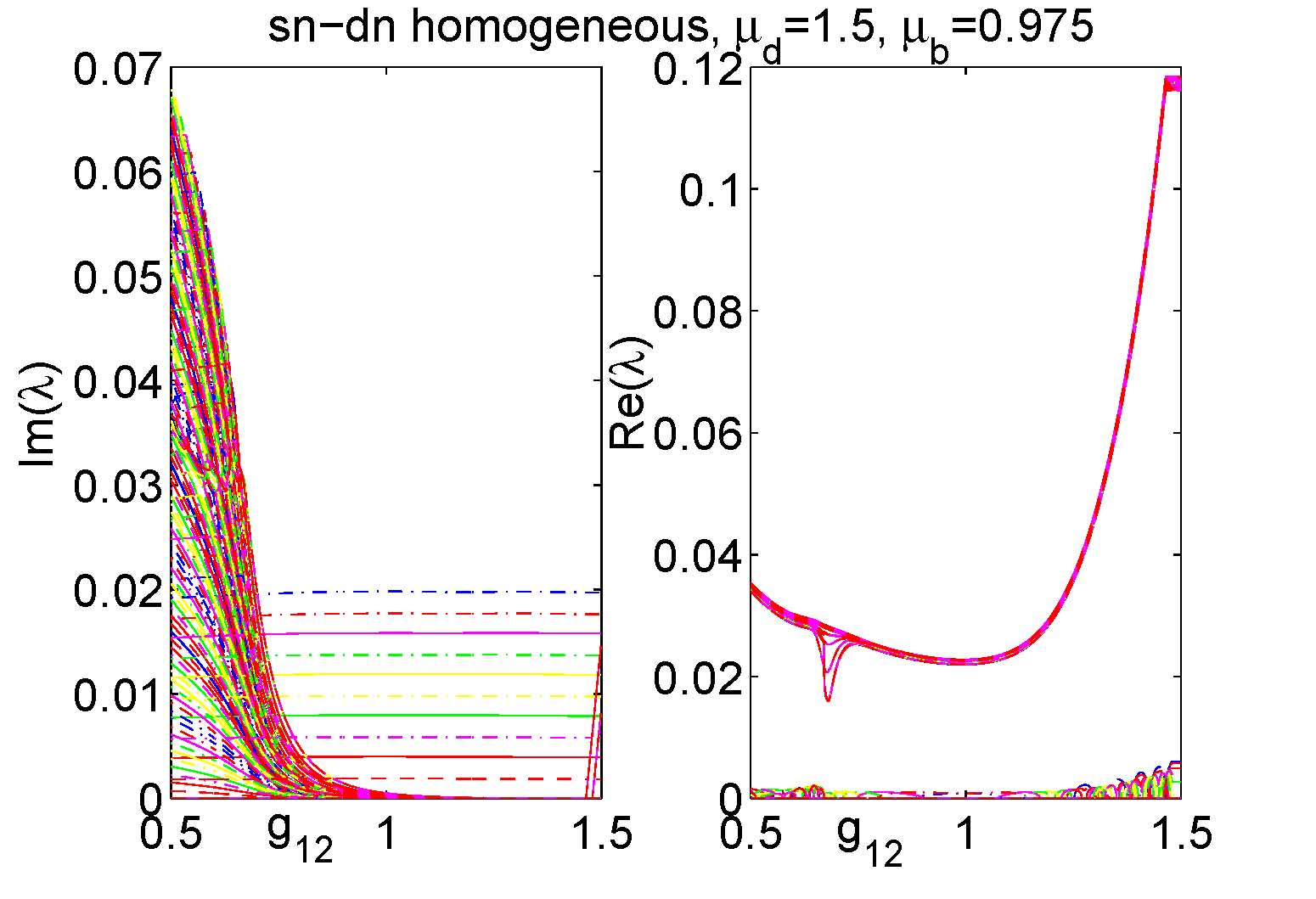}  
\end{tabular}
\caption{The left panel shows the spectrum of the sn-dn periodic solution
for $\mu_d=1.5$ and $\mu_b=0.975$ as a function of $g_{12}$ using the
finite difference method. The right panel once again shows the same
spectrum but with the Hill's method (for 11 values of the relevant
angle $\theta$) incorporated in the computation. The relevant waveform
is generically unstable, although it is most weakly so on the slightly
immiscible side.}
\label{sn_dn_homogeneous_spectrum_g12_vary}
\end{figure}

Similar results, still without a trap (i.e., in the homogeneous BEC
realm) are shown for the lattice solution where the bright solitons
are in phase (the sn-dn lattice) in Fig.~\ref{sn_dn_homogeneous_stationary}.
This solution is also found to exist for more general conditions than
the ones for which it is traced analytically earlier. Here, we fix
$\mu_d=1.5$ and $\mu_b=0.975$ and again vary $g_{12}$. Again a variation
is discernible as the miscibility-immiscibility threshold is
traversed to wider bright solitary waves, while on the immiscible
side these are well separated and far narrower. The stability is again
computed with the two methods (finite difference method for the
linearization eigenvalue computation and also
its variant incorporating the Hill's approach). As is shown
in Fig.~\ref{sn_dn_homogeneous_spectrum_g12_vary}, once again there
appears a minimal growth rate (and hence a maximal life time
of the pertinent waveforms) to be applicable weakly on the
immiscible side (yet fairly closely to the miscibility-immiscibility
threshold). As one proceeds deeper on the immiscible or for that
matter on the miscible side, the solutions become more strongly
unstable and hence less likely to be observable even transiently.

%
% sn-dn homogeneous: FDM
%

\begin{figure}
\begin{tabular}{cc}
\includegraphics[width=8cm]{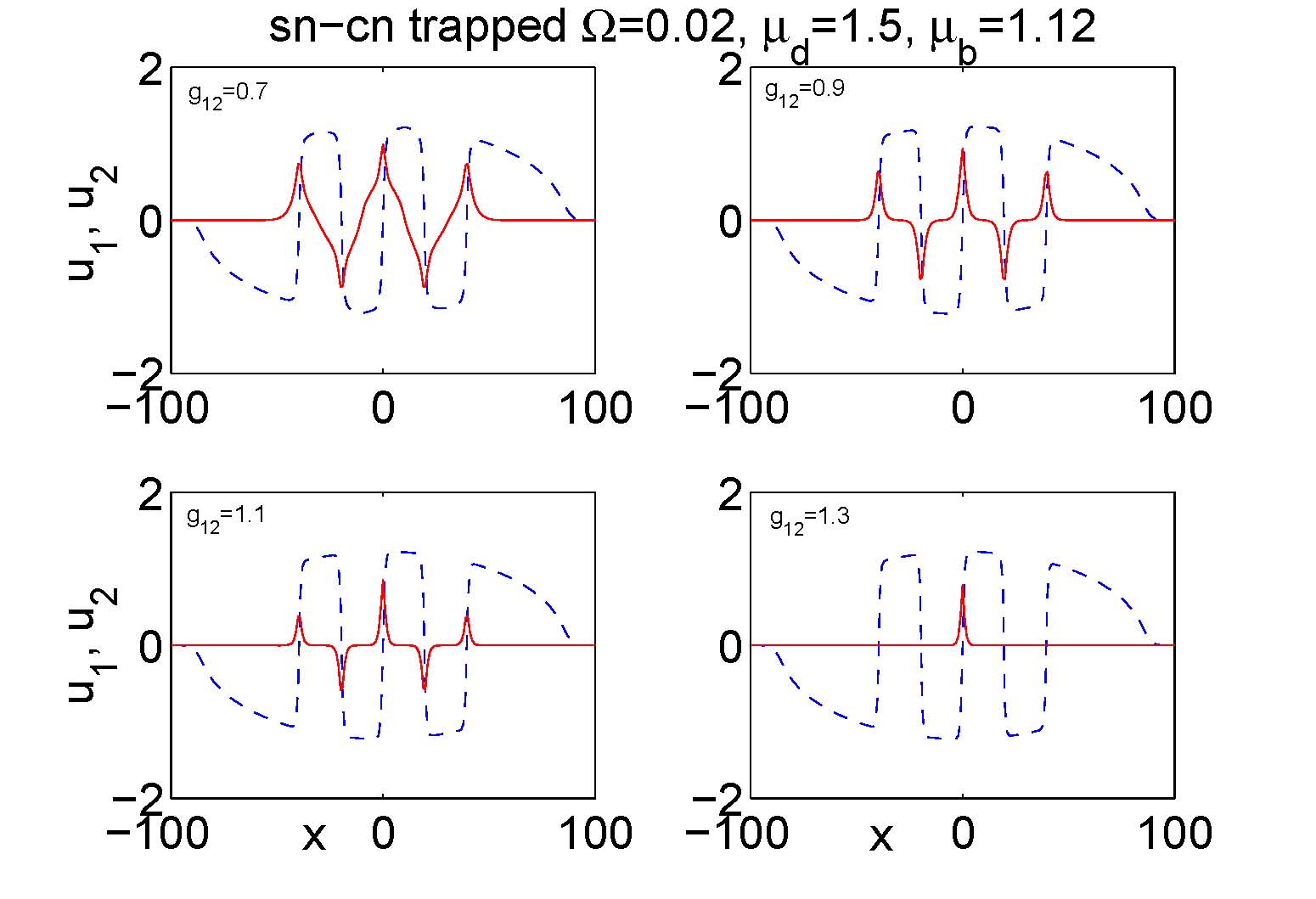}
\includegraphics[width=8cm]{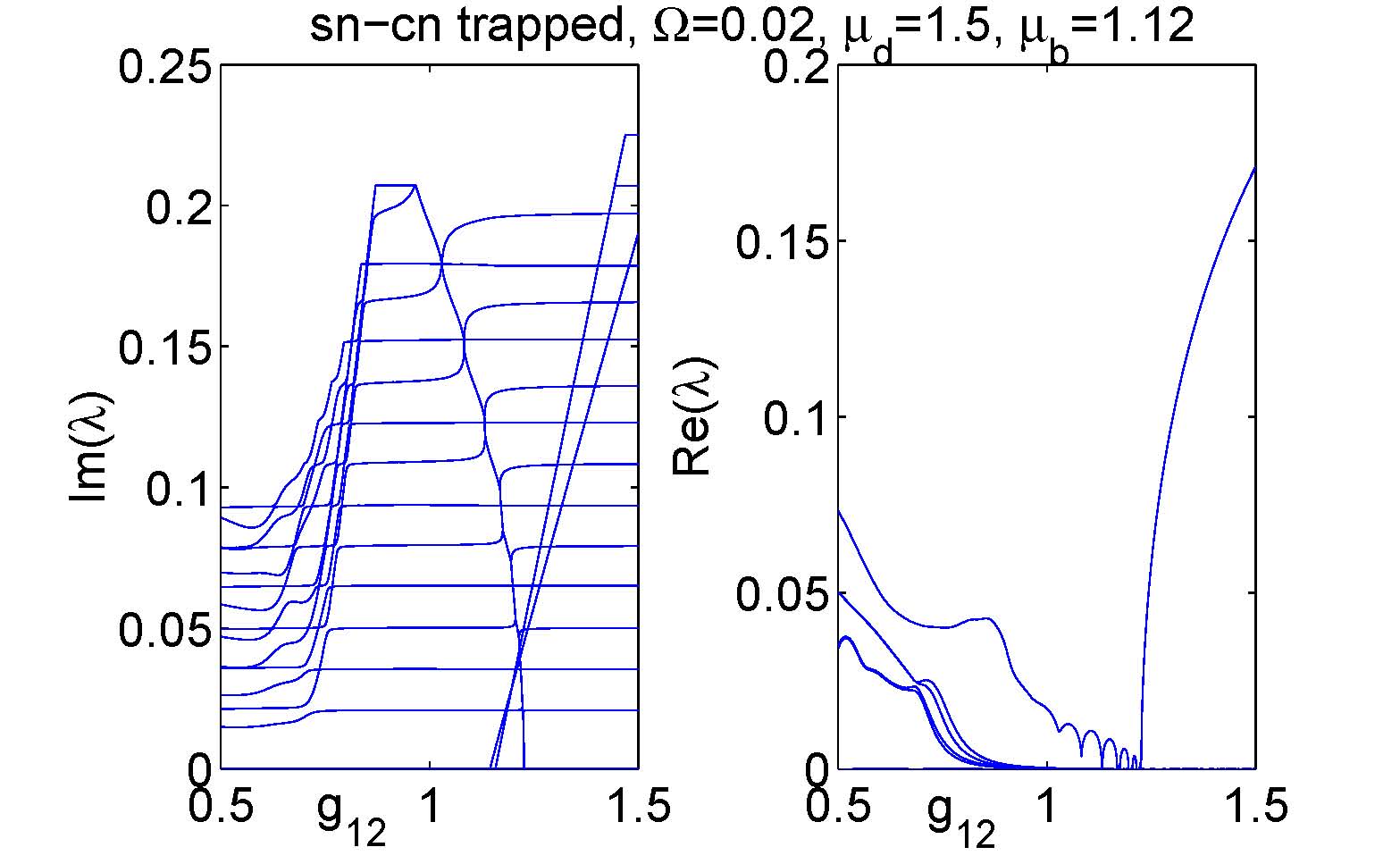}
\end{tabular}
\caption{The left panel of the figure shows the stationary profile of trapped sn-cn type solutions for $g_{12}=0.7, 0.9, 1.1, 1.3$
on the top left, top right, bottom left and bottom right panel respectively.
The trapping  frequency is $\Omega=0.02$, while the
chemical potentials are $\mu_d=1.5$ and $\mu_b=1.12$. When $g_{12}$ is
about 1.2, it is interesting to note that the combination of
the trap and the immiscibility only permits to one of the bright
peaks (the central one) to persist, while the rest have disappeared.
The right panel shows the linearization spectrum (again, imaginary
and real parts) as a function of $g_{12}$.}
\label{sn_cn_trapped_stationary}
\end{figure}

\begin{figure}
\begin{tabular}{cc}
\includegraphics[width=8cm]{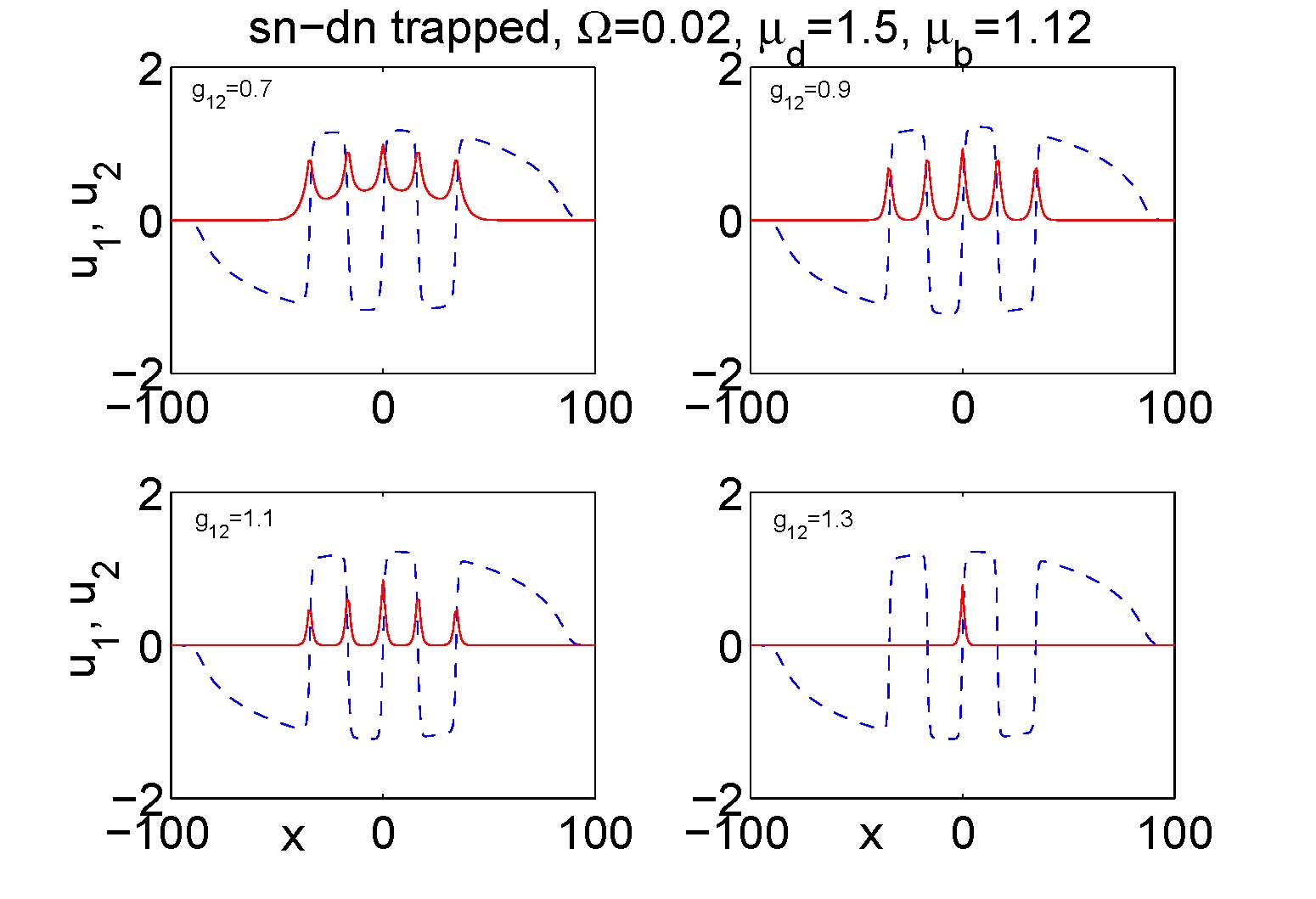}
\includegraphics[width=8cm]{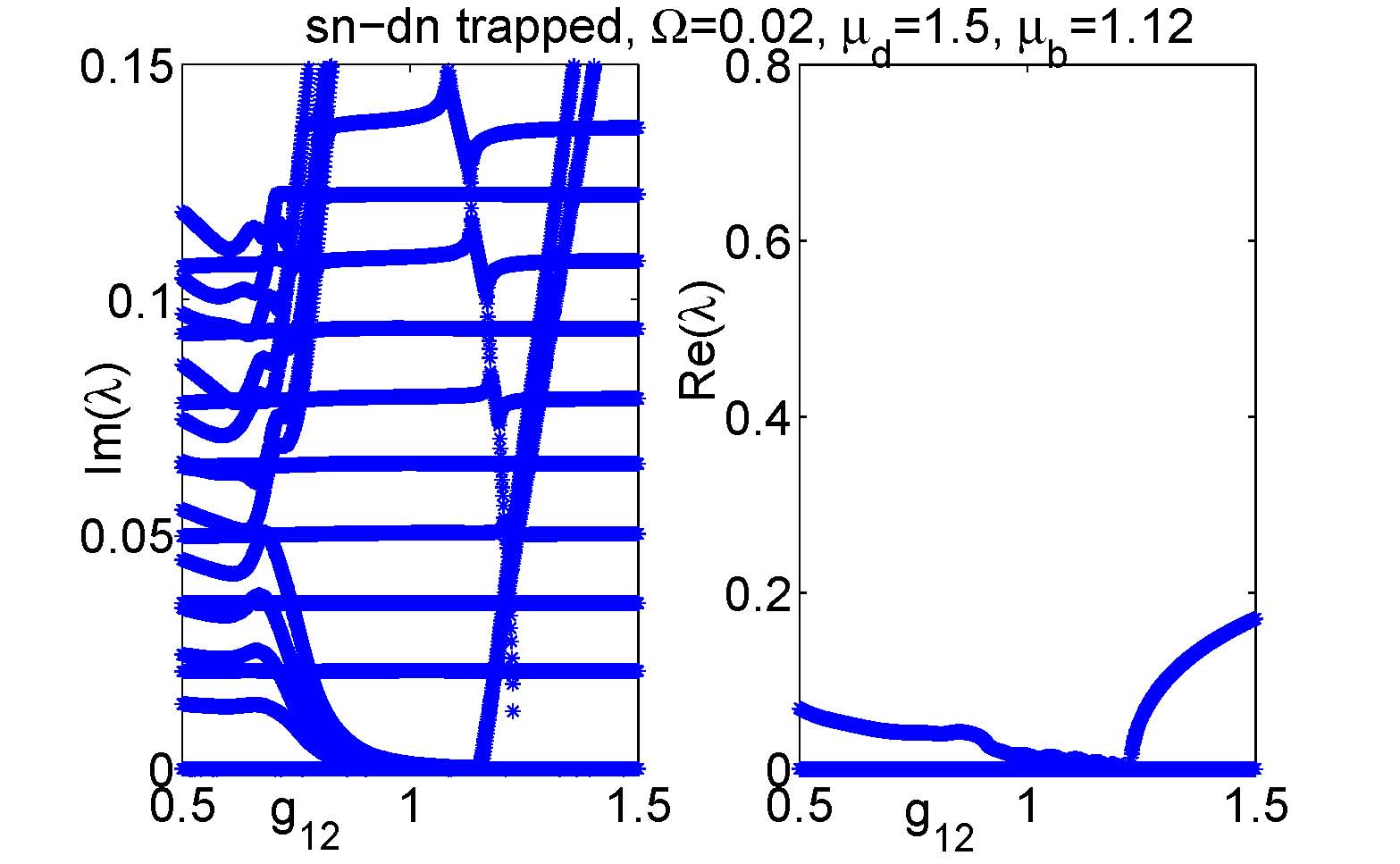}
\end{tabular}
\caption{The left panel shows the stationary profile of sn-dn type
solutions in the presence of a trap, for $g_{12}=0.7, 0.9, 1.1, 1.3$
on the top left, top right, bottom left and bottom right panel respectively.
The trap frequency is $\Omega=0.02$ and
the chemical potentials are $\mu_d=1.5$ and $\mu_b=1.12$.
The right panel again shows the corresponding linearization
eigenvalues as a function $g_{12}$.}
%When $g_{12}$ is larger than 1.2, the dn solutions do
%not have 5 parts any more.}
\label{sn_dn_trapped_stationary}
\end{figure}

Finally, we now turn to the consideration of trapped variants of
the lattice solutions, as an extension of both the single DB
trapped solution, but also the homogeneous BEC lattices of
sn-cn and sn-dn waveforms. Our numerical computations for the
two types of lattices are shown, respectively, in
Figs.~\ref{sn_cn_trapped_stationary} and~\ref{sn_dn_trapped_stationary}.
In Fig.~\ref{sn_cn_trapped_stationary},
we can observe the persistence of the sn-cn
lattice in the presence of the trap,
although an intriguing by-product of the interplay between the presence
of a finite $\Omega \neq 0$ and a progressively stronger inter-species
interaction ${g}_{12}$ is the gradual depletion of the outer
bright peaks, eventually (see bottom right for ${g}_{12}=1.3$)
in favor of a single peak at the center. The stability results again
illustrate that even in the presence of the trap the instability
growth rates of the solution are again minimal in the vicinity of
the miscibility-immiscibility threshold (although in this case,
the absolute minimum of the growth rates appears to
be shifted towards the weakly miscible side). Fairly similar
conclusions, both as regard the ``squeezing'' (and eventual elimination)
of the bright peaks, as well as the minimal growth rates on the
weakly miscible side can be observed also for the trapped
variant of the {\rm sn}-{\rm dn} solution in Fig.~\ref{sn_dn_trapped_stationary}.
%tilde
%

\section{Future Challenges \& Conclusions}

In the present work, we have revisited the theme of dark-bright
solitary waves in atomic Bose-Einstein condensates. We have considered
such nonlinear structures in the presence of general interaction
coefficients, motivated by the tunability of the scattering lengths,
by means of Feshbach resonances which, in turn, permit a tunability
of the intra- and inter-species effective nonlinear interaction coefficients.
 We have seen that remarkably the DB states in the presence and absence
of the trap persist for a very broad range of inter-species interactions
(this has been our principal control parameter). Within a suitably narrow range,
we have been able to predict such a variation even analytically.
We have also analytically predicted the motion of these DB solitary
waves, identifying it as a harmonic oscillation within a parabolic trap.
However, we have also gone well beyond individual dark-bright solitary waves,
and have explored extended variants thereof, in the form of DB soliton
lattices. Such lattices were even predicted analytically in the form
of cnoidal wave solutions with the bright components forming
adjacent in-phase or out-of-phase pairs, i.e., sn-dn and sn-cn
solutions, respectively. While these solutions were found in the
homogeneous BEC, remarkable it was possible to computationally
extend them even in the trapped case. Finally, their stability
was also numerically explored, finding that they can be least
unstable in the vicinity of miscibility-immiscibility threshold.

Given the extensive level of control of recent experiments on
multi-component, DB-soliton-bearing experiments (see, for
instance,~\cite{becker,pe1,pe2,pe2a,pe3,pe4,pejc}) and the
ability to tune scattering lengths by means of the Feshbach
resonance mechanism \cite{FR2}, we believe that the type of states/configurations
proposed herein should be well within experimental reach.
Additionally, it would be extremely interesting to generalize
relevant configurations in higher dimensions. So far, to the best
of our knowledge, only configurations of a single
%~\cite{kody_prl}
or two~\cite{kody_prl,martina2} vortex-bright states have been proposed
and the pertinent understanding of their dynamics is purely numerical.
Obtaining an analytical description of their motion and generalizing
such states in the realm of lattices would be a particularly interesting
possibility in its own right, in a way perhaps reminiscent of other
types of multi-component lattices (of vortex molecules) such as the
ones proposed in Ref.~\cite{ueda}. Relevant studies are currently
in progress and will be reported in future publications.

\section*{Appendix: Finite Difference, Finite Difference with Hill and Hill's method}
In order to determine the linear stability of the stationary solution $(u_{1,0}, u_{2,0})$, we assume a general perturbation
around it in the form
\begin{eqnarray}
u_d &=& u_{1,0}+\epsilon\left(a(x)e^{\lambda t}+b(x)^{*}e^{\lambda^{*}t}\right)
\label{perturb1}
\\
u_b &=& u_{2,0}+\epsilon\left(c(x)e^{\lambda t}+d(x)^{*}e^{\lambda^{*}t}\right)
\label{perturb2}
\end{eqnarray}
and substitute in the dynamical equations, computing only
the $O(\epsilon)$ corrections.
The relevant linear eigenvalue problem is then written as
%
%\begin{eqnarray}
%\lambda
%\end{eqnarray}
%
%
\[
\lambda \left(\begin{array}{c}
a \\
b \\
c \\
d \end{array} \right)=\left( \begin{array}{cccc}
A_{11} & A_{12} & A_{13} & A_{14} \\
A_{21} & A_{22} & A_{23} & A_{24} \\
A_{31} & A_{32} & A_{33} & A_{34} \\
A_{41} & A_{42} & A_{43} & A_{44} \end{array} \right)
\left(\begin{array}{c}
a \\
b \\
c \\
d \end{array} \right)
\],
where $\lambda$, $(a,b,c,d)$ are the eigenvalues and eigenvectors, respectively. In particular, the matrix elements are:
\begin{eqnarray}
A_{11} &=& -\frac{1}{2}\partial_{xx}-\mu_d+V(x)+2g_{11}|u_{1,0}|^2 \nonumber \\
&+& g_{12}|u_{2,0}|^2
\\
A_{22} &=& - A_{11}
\\
A_{33} &=& -\frac{1}{2}\partial_{xx}-\mu_b+V(x)+2g_{22}|u_{2,0}|^2 \nonumber \\
&+& g_{12}|u_{1,0}|^2
\\
A_{44} &=& - A_{33}
\\
A_{12} &=& g_{11}u_{1,0}^2
\\
A_{13} &=& g_{12}u_{1,0}u_{2,0}^{*}
\\
A_{14} &=& g_{12}u_{1,0}u_{2,0}
\\
A_{21} &=& -A_{12}^*
\\
A_{23} &=& -A_{14}^*
\\
A_{24} &=& -A_{13}^*
\\
A_{31} &=& A_{13}^*
\\
A_{32} &=& A_{14}
\\
A_{34} &=& g_{22}u^2_{2,0}
\\
A_{41} &=& -A_{22}^*
\\
A_{42} &=& -A_{32}
\\
A_{43} &=& -A_{34}.
\end{eqnarray}
Now we briefly discuss two  methods for studying
the above linear eigenvalue problem. For the finite difference method,
we discretize the eigenvector and the Jacobian matrix, i.e.,
work with the grid $x_n=x_1+(n-1)\Delta x$.
For the eigenvectors $(a,b,c,d)$, we then have $a(x)=(a(x_1),a(x_2),\cdots,a(x_n))$, $b(x)=(b(x_1),b(x_2),\cdots,b_(x_n))$, $c(x)=(c(x_1),c(x_2),\cdots,c(x_n))$ and $d(x)=(d(x_1),d(x_2),\cdots,d(x_n))$.
The resulting matrix eigenvalue-eigenvector problem can thus be numerically
solved. \\
For the finite difference method with Hill's method incorporated~\cite{hills},
we
select a number of values for
$\theta\in [0,2\pi)$, and make the following changes based on the finite difference method

\begin{eqnarray}
A_{11}(1,n)\rightarrow A_{11}(1,n)e^{i\theta}
\\
A_{22}(1,n)\rightarrow A_{22}(1,n)e^{i\theta}
\\
A_{33}(1,n)\rightarrow A_{33}(1,n)e^{i\theta}
\\
A_{44}(1,n)\rightarrow A_{44}(1,n)e^{i\theta}
\\
A_{11}(n,1)\rightarrow A_{11}(n,1)e^{-i\theta}
\\
A_{22}(n,1)\rightarrow A_{22}(n,1)e^{-i\theta}
\\
A_{33}(n,1)\rightarrow A_{33}(n,1)e^{-i\theta}
\\
A_{44}(n,1)\rightarrow A_{44}(n,1)e^{-i\theta}.
\end{eqnarray}
Then we evaluate the eigenvalues and eigenvectors of the matrix
$A$ over a period of the periodic solution of interest and
superpose the relevant spectra obtained for different values of
$\theta$.

In the present work, we computed the spectrum with
finite differences and finite differences incorporating Hill's method
(over a period) and confirmed the agreement between the two.
One can alternatively also consider the direct Hill's method as
described e.g. in~\cite{hills}.

%\\
%For now, we just incorporated the numerical results for the finite difference and finite difference with Hill's method.

%In
%Fig.~\ref{sn_cn_homogeneous_stationary} to Fig.~\ref{sn_cn_homogeneous_spectrum%_g12_vary_fdm_hill}, they show the stationary
%solutions, the spectrum by finite difference, and spectrum by finite difference with Hill respectively for sn-cn type of solutions
%in the homogeneous case.
%Similarly, in Fig.~\ref{sn_dn_homogeneous_stationary} to Fig.~\ref{sn_dn_homogeneous_spectrum_g12_vary_fdm_hill} are for the sn-dn
%type of solutions. For the trapped cases, Fig.~\ref{sn_cn_trapped_stationary} t%o Fig.~\ref{sn_cn_trapped_spectrum_g12_vary_fdm_hill}
%and Fig.~\ref{sn_dn_trapped_stationary} to Fig.~\ref{sn_dn_trapped_spectrum_g12%_vary_fdm_hill} demonstrate the sn-cn and sn-dn solutions
%respectively.


\begin{thebibliography}{99}

\bibitem{manakov} S. V. Manakov, Sov. Phys. JETP {\bf 38}, 248 (1974).

\bibitem{APT} M. J. Ablowitz, B. Prinari, and A. D. Trubatch,
{\it Discrete and Continuous Nonlinear Schrödinger Systems},
Cambridge University Press (Cambridge, 2004).


\bibitem{buschanglin} Th. Busch and J. R. Anglin,
Phys. Rev. Lett. {\bf 87}, 010401 (2001).


\bibitem{becker} C. Becker, S. Stellmer, P. Soltan-Panahi, S.
D{\"o}rscher, M. Baumert, E.-M. Richter, J. Kronj{\"a}ger,
K. Bongs, and K. Sengstock, Nature Phys. {\bf 4}, 496 (2008).


\bibitem{pe1} S. Middelkamp, J. J. Chang, C. Hamner, R. Carretero-Gonz{\'a}lez, P. G. Kevrekidis, V. Achilleos, D. J. Frantzeskakis, P. Schmelcher, and P. Engels, Phys. Lett. A {\bf 375}, 642 (2011).


\bibitem{pe2}  C. Hamner, J. J. Chang, P. Engels, and M. A. Hoefer,
Phys. Rev. Lett. {\bf 106}, 065302 (2011).


\bibitem{pe2a} D. Yan, J. J. Chang, C. Hamner, P. G. Kevrekidis, P. Engels,
V. Achilleos, D. J. Frantzeskakis, R. Carretero-Gonz{\'a}lez, and P. Schmelcher,
Phys. Rev. A {\bf 84}, 053630 (2011).

\bibitem{vpg1} C. Y. Yin, N. G. Berloff, V. M. P{\'e}rez-Garc{\'i}a,
D. Novoa, A. V. Carpentier and H. Michinel,
Phys. Rev. A {\bf 83}, 051605 (2011).

\bibitem{vpg2} V. A. Brazhnyi and V. M. P{\'e}rez-Garc{\'i}a,
Chaos, Solitons and Fractals, {\bf 44}, 381 (2011).

\bibitem{pe3} M. A. Hoefer, J. J. Chang, C. Hamner, and P. Engels,
Phys. Rev. A {\bf 84}, 041605 (2011).


\bibitem{pe4} D. Yan, J. J. Chang, C. Hamner, M. Hoefer, P. G. Kevrekidis, P. Engels, V. Achilleos,
D. J. Frantzeskakis, and J. Cuevas, J. Phys. B: At. Mol. Opt. Phys. {\bf 45}, 115301 (2012).
%arXiv:1202.2777.

\bibitem{pejc} A. \'{A}lvarez, J. Cuevas, F. R. Romero, C. Hamner, J. J. Chang, P. Engels, P. G. Kevrekidis
and D. J. Frantzeskakis, J. Phys. B  At. Mol. Opt. Phys. {\bf 46},  065302 (2013).

\bibitem{jan_stock} J. Stockhofe, P.G. Kevrekidis, D.J. Frantzeskakis
and P. Schmelcher, J. Phys. B  At. Mol. Opt. Phys. {\bf 44}, 191003 (2011).

\bibitem{kody_prl}  J.J. Garc{\'i}a-Ripoll and V.M. P{\'e}rez-Garc{\'i}a
Phys. Rev. Lett. {\bf 84}, 4264 (2000).

\bibitem{martina2} K. J. H. Law, P. G. Kevrekidis, and L. S. Tuckerman,
Phys. Rev. Lett. {\bf 105}, 160405 (2010);
M. Pola, J. Stockhofe, P. Schmelcher, and P. G. Kevrekidis,
Phys. Rev. A {\bf 86}, 053601 (2012).

\bibitem{FR} S. Inouye, M. R. Andrews, J. Stenger, H.-J. Miesner D. M.
Stamper-Kurn, and W. Ketterle, Nature (London) \textbf{392}, 151 (1998); J.
L. Roberts, N. R. Claussen, J. P. Burke, Jr., C. H. Greene, E. A. Cornell,
and C. E. Wieman, Phys. Rev. Lett. 81, 5109 (1998); E. A. Donley, N. R.
Claussen, S. L. Cornish, J. L. Roberts, E. A. Cornell, and C. E. Wieman,
Nature (London) \textbf{412}, 295 (2001).

\bibitem{FR2} G. Thalhammer, G. Barontini, L. De Sarlo, J. Catani, F. Minardi, and M. Inguscio,
Phys. Rev. Lett. {\bf 100}, 210402 (2008); S. B. Papp, J. M. Pino and C. E. Wieman,
Phys. Rev. Lett. {\bf 101}, 040402 (2008).

\bibitem{csire} G. Csire, D. Schumayer, and B. Apagyi,
Phys. Rev. A {\bf 82}, 063608 (2010).

\bibitem{ieee} V. V. Afanasjev, E. M. Dianov and V. N. Serkin,
IEEE J. Quantum Electron. {\bf 25}, 2656 (1989).

\bibitem{uzunov} N. A Kostov and I. M. Uzunov, Opt. Commun. {\bf 89}, 389 (1992).

\bibitem{vas} V. Achilleos, P. G. Kevrekidis, V. M. Rothos, and D. J. Frantzeskakis,
Phys. Rev. A {\bf 84}, 053626 (2011); V. Achilleos, D. Yan, P. G. Kevrekidis, and D. J. Frantzeskakis,
New J. Phys. {\bf 14}, 055006 (2012).


\bibitem{revfr} D. J. Frantzeskakis, J. Phys. A: Math. Theor. {\bf 43}, 213001 (2010).


\bibitem{emergent} P. G. Kevrekidis, D. J. Frantzeskakis, and R.
Carretero-Gonz{\'a}lez, {\it Emergent Nonlinear Phenomena in
Bose-Einstein Condensates}, Springer-Verlag (Berlin, 2008).

\bibitem{hills} B. Deconinck and J. N. Kutz,
J. Comp. Physics {\bf 219}, 296 (2006).

\bibitem{ueda} K. Kasamatsu, M. Tsubota, and M. Ueda,
Phys. Rev. Lett. {\bf 93}, 250406 (2004).


\end{thebibliography}
\end{document}